\documentclass[%
aip,
amsmath,amssymb,
reprint,%
]{revtex4-1}
\usepackage{natbib}
\usepackage{graphicx}
\usepackage{dcolumn}
\usepackage{bm}

\usepackage[utf8]{inputenc}
\usepackage[T1]{fontenc}
\usepackage{mathptmx}
\usepackage{mathptmx}
\usepackage{siunitx}
\usepackage{upgreek}

\begin{document}

\preprint{}

\title{Shock-wave-induced cavitation of silicone oils} 



\author{Justin Huneault}
\affiliation{Department of Mechanical Engineering, McGill University, 817 Sherbrooke St. West, Montreal, QC H3A 0C3, Canada}

\author{Andrew Higgins}
\email[]{andrew.higgins@mcgill.ca}
\affiliation{Department of Mechanical Engineering, McGill University, 817 Sherbrooke St. West, Montreal, QC H3A 0C3, Canada}


\date{\today}

\begin{abstract}
The cavitation threshold of polydimethylsiloxane (silicone) oils was studied using the planar impact of flyer plates to generate large transient negative pressures within the liquids. The plate-impact experiments used a 64-mm-bore gas-gun to launch thin sabot-supported flyer plates onto liquid capsule targets in which a thin Mylar diaphragm formed a free surface at the back of the sample. The shock wave driven into the target capsule by the flyer impact placed the silicone oil in tension upon reflection from the rear free surface, eventually causing the sample to cavitate. The spall strength, or critical tension which cavitates the liquid, was determined by monitoring the free-surface velocity using a photonic Doppler velocimetry system. This study explored the effect of viscosity and loading strain rate on a system of three silicone oils having vastly different viscosities (\SIrange{4.8e-2}{2.9e1}{\pascal\second}), but otherwise similar properties. The spall strength was found to remain constant over the ranges of strain rate and viscosities probed in this work. A comparison of the experimental results to models for the cavitation threshold of liquids suggested that homogeneous nucleation of bubbles was the dominant mechanism for tension relief at the onset of cavitation.
\end{abstract}


\maketitle 

\section{INTRODUCTION}
\label{introduction}
Liquids can be made to support significant tension despite the fact that they are in a non-equilibrium state at pressures below their vapor pressure.~\cite{Trevena1987} Tension in the liquid is relieved through the nucleation and growth of vapor cavities in a process known as cavitation. Surface tension creates an energy barrier which must be overcome for cavitation bubbles growth to occur~\cite{Fisher1948}, thus allowing liquids to exist in a metastable state of tension equivalent to that of a superheated fluid. Cavitation bubbles typically form from heterogeneous nucleation sites, including free bubbles and bubbles attached to impurities within the liquid or at container walls.~\cite{Kedrinskii2005} However, for sufficiently pure liquids or in cases where tension is applied to the liquid on a sufficiently short timescale, cavitation bubbles may nucleate from random molecular fluctuations in a process of homogeneous nucleation.~\cite{Fisher1948,Zeldovich1992,Carlson1975,Utkin2010} 

The reflection of a shock wave from a free surface can momentarily place a liquid in a state of tension due to the interaction of the resulting reflected expansion front with an expansion front behind the shock. This highly dynamic phenomenon causes the liquid to be stretched until sufficient tension is reached to cause rupture through the formation and growth of bubbles. The reflection of a pressure pulse from a free surface has been used extensively to study the behavior of liquids in tension~\cite{Bull1956,Erlich1971,Carlson1975,Trevena1987,Utkin2010}, and plays an important role in a number of physical processes, including underwater explosions~\cite{Kolsky1949,Cole1948,Kedrinskii2005}, the fracture of kidney stones using shock wave lithotripsy~\cite{Zhu2002}, and the formation of jet-like instabilities during the explosively driven dispersal of liquids.~\cite{Milne2017} Shock-wave-induced cavitation is also of interest for a recently proposed Magnetized Target Fusion concept in which a spherically imploding shock wave is used to collapse a liquid metal cavity onto a plasma target in order to reach fusion conditions.~\cite{Laberge2008,Laberge2009,Suponitsky2014, Suponitsky2017} 

Bull~\cite{Bull1956} was the first to develop an experimental apparatus that used pressure wave reflection from a free surface to measure the maximum tension in the liquid prior to cavitation. In these \textit{bullet-piston} experiments, an impact-driven pressure pulse is fed into a long vertical column of liquid with a free surface in contact with a gas atmosphere near the top.~\cite{Bull1956} The incident pressure pulse reflects from the free surface as an expansion front that generates significant transient tension in the liquid. Pressure sensors mounted along the liquid column measure the incident pressure pulse that moves up the column as well as the peak tension in the liquid caused by the reflected tensile pulse.~\cite{Trevena1987} To access even greater loading rates, Erlich et al.~\cite{Erlich1971} pioneered the use of spall experiments, which had previously been used to study the dynamic fracture of solids, to study cavitating liquids. In these experiments a planar shock wave is driven into a target liquid, typically using the impact of a flyer plate, and allowed to reflect from a free surface at the rear of the sample. As will be described in Section~\ref{PIexperiments}, the maximum tension within the liquid can be inferred by observing the time evolution of the free-surface velocity.~\cite{Antoun2003} Such experiments determine the critical tension at which the liquid begins to fail in tension under the loading conditions of the experiment, a value typically referred to as \textit{spall strength}. In spall tests, the loading dynamics ensure that cavitation originates within the bulk liquid, rather than at the walls of the container, which means that the maximum tension is a true measurement of the limit of cohesion of the liquid.~\cite{Utkin2010} 

A number of studies have looked at the effect of viscosity and strain rate on the cavitation threshold of liquids. Using a bullet-piston type experiment, Bull~\cite{Bull1956} found an empirical power law relationship ($P_\mathrm{s}\sim \eta^{0.2}$) between dynamic viscosity ($\eta$) and the tension at the onset of cavitation ($P_\mathrm{s}$) for an assortment of liquids spanning  four orders of magnitude in viscosity. The observed increase in the cavitation threshold with increasing viscosity was attributed to a larger resistance to viscous growth of voids.~\cite{Bull1956} Couzens and Trevena~\cite{Couzens1974} studied the relationship between spall strength and viscosity for silicone oils  over a range of three orders of magnitude using bullet-piston experiments and found a similar, albeit less sensitive relationship between cavitation threshold and viscosity ($P_\mathrm{s}\sim \eta^{0.06}$). A comparison of critical tension measurements for glycerol obtained from the bullet-piston experiments of Bull~\cite{Bull1956b} and the spall experiments of Erlich et al.~\cite{Erlich1971}, shows that the critical tension was seen to increase from approximately \SI{6}{\mega\pascal} at bullet-piston strain rates ($\dot{\epsilon}\approx$\SI{e2}{\per\second}) to \SI{23}{\mega\pascal} at spall experiment strain rates ($\dot{\epsilon}\approx$\SI{e5}{\per\second}). Grady developed an energy-based spall strength model involving the viscous growth of pre-existing bubbles in a liquid~\cite{Grady1988}, and showed relatively good agreement with the data of Erlich et al. and Bull, indicating that the observed increase in cavitation threshold with increasing strain rate may be attributed to the greater viscous dissipation as the deformation rate is increased.

In contrast to the results quoted above, which indicated that the cavitation threshold of liquids may be determined by a heterogeneous nucleation mechanism, previous studies performed at high strain rates ($\dot{\epsilon}>$\SI{e4}{\per\second}) have shown evidence that the spall strength of liquids is determined primarily by a homogenous cavitation mechanism. Carlson and Levine~\cite{Carlson1975} used an electron beam to determine the spall strength of glycerol over a wide range of temperatures (491~to~623~K), where the viscosity changes by five orders of magnitude. The experiments showed a factor of eight decrease in spall strength as temperature was increased.~\cite{Carlson1975} Utkin and Sosikov~\cite{Utkin2010} used flyer-driven spall experiments to compare the effect of varying strain rate on the spall strength of liquids having vastly different viscosities: glycerol, hexane, methanol, and water. They observed that the spall strength in hexane, methanol, and water was nearly independent of strain rate over a range of approximately \SIrange{e4}{e5}{\per\second}~\cite{Utkin2010}, which was attributed to the weak dependence of the cavitation threshold on the loading rate according to homogeneous nucleation theory.~\cite{Fisher1948,Zeldovich1992} In contrast, the spall strength of the glycerol samples was seen to increase by a factor of 2.5 over a variation in strain rate of \SIrange{1.5e4}{2.0e5}{\per\second}. Both Carlson~\cite{Carlson1975} and Utkin~\cite{Utkin2010} used models to show that the observed variation in the spall strength of glycerol with temperature and strain rate can be attributed to the fact that the loading rates in the experiments were on the same timescale as the relaxation time needed to reach the steady-state homogeneous void nucleation rate, rather than being caused by a heterogeneous viscous void growth mechanism. They claimed that for nearly pure liquids placed in tension at sufficiently high strain rates, the dominant mechanism which relieves tension and determines the spall strength of the liquid is the homogeneous nucleation of bubbles.~\cite{Carlson1975, Utkin2010} From the theoretical work of Fisher~\cite{Fisher1948} and Zeldovich~\cite{Zeldovich1992}, the critical tension required to cavitate a liquid should be strongly affected by its surface tension, but only weakly related to its viscosity or the loading rate of the experiment, provided that the relaxation time needed to reach the steady-state void nucleation rate is much smaller than the loading rate.~\cite{Carlson1975}

This paper, which builds upon the results of a previously reported preliminary study~\cite{Huneault2018}, will examine the effect of viscosity and strain rate on the spall strength of a set of silicone oils which have vastly different viscosities, but otherwise similar material properties. The experiments will focus on observing the relationship between the measured spall strength and the liquid properties or loading characteristics. The results will be compared to existing models for the limiting tension of liquids, in order to offer insight into the mechanism of cavitation nucleation.

\section{MATERIALS AND METHODS}
\label{mandm}
\subsection{Materials}
\label{materials}
This study will examine the spall behavior of a set of three polydimethylsiloxane (PDMS) silicone oils, sourced from Clearco Products Co., with dynamic viscosities ($\eta$) that span over two orders of magnitude, but otherwise similar material properties (density ($\rho$), speed of sound ($c_0$), surface tension ($\sigma$), and bulk modulus ($K$)). The relevant material properties for the three fluids can be seen in Tab.~\ref{table1}, where the oils have been identified by their nominal kinematic viscosity in Stokes (St). As can be seen from the molecular weight ($M$) values in Tab.~\ref{table1}, the difference in viscosity between the oils is due to a variation in the average length of the polymer chains, which affects the resistance of the fluid to shear.~\cite{Patterson1998} 

\begin{table*}
\caption{Properties of the silicone oils used in the study.}
\label{table1}
\begin{ruledtabular}
\begin{tabular}{lccccccc}
&  & $\eta$\footnotemark[1] & $\rho_0$\footnotemark[1] & $c_0$\footnotemark[2] & $\sigma$\footnotemark[1] & $K$\footnotemark[3] & $M$\footnotemark[1] \\
Silicone Oil & Product No. & (\SI{}{\pascal\second}) & (\SI{}{\kilo\gram\per\metre\cubed}) & (\SI{}{\metre\per\second}) & (\SI{}{\newton\per\metre}) & (\SI{}{\giga\pascal}) & (\SI{}{\kilo\gram\per\mole})\\
\hline\noalign{\smallskip}
0.5 St  & PSF-50cSt       & \num{4.80e-2} & 960 & 1004 & \num{2.08e-2} & 0.97 & \num{3.78e0} \\
10 St   & PSF-1,000cSt   & \num{9.71e-1} & 971 & 1004 & \num{2.12e-2} & 0.98 & \num{2.80e1} \\
300 St & PSF-30,000cSt & \num{2.93e1}  & 976 & 1004 & \num{2.13e-2} & 0.98 & \num{9.17e2} \\
\end{tabular}
\end{ruledtabular}
\footnotetext[1]{Taken from manufacturer data.}
\footnotetext[2]{Taken from a study using a similar \ silicone oil.\cite{Leibacher2015}}
\footnotetext[3]{Calculated from $K=c_0^2\rho_0$.}
\end{table*}

\begin{figure}
\includegraphics[width=0.9\columnwidth]{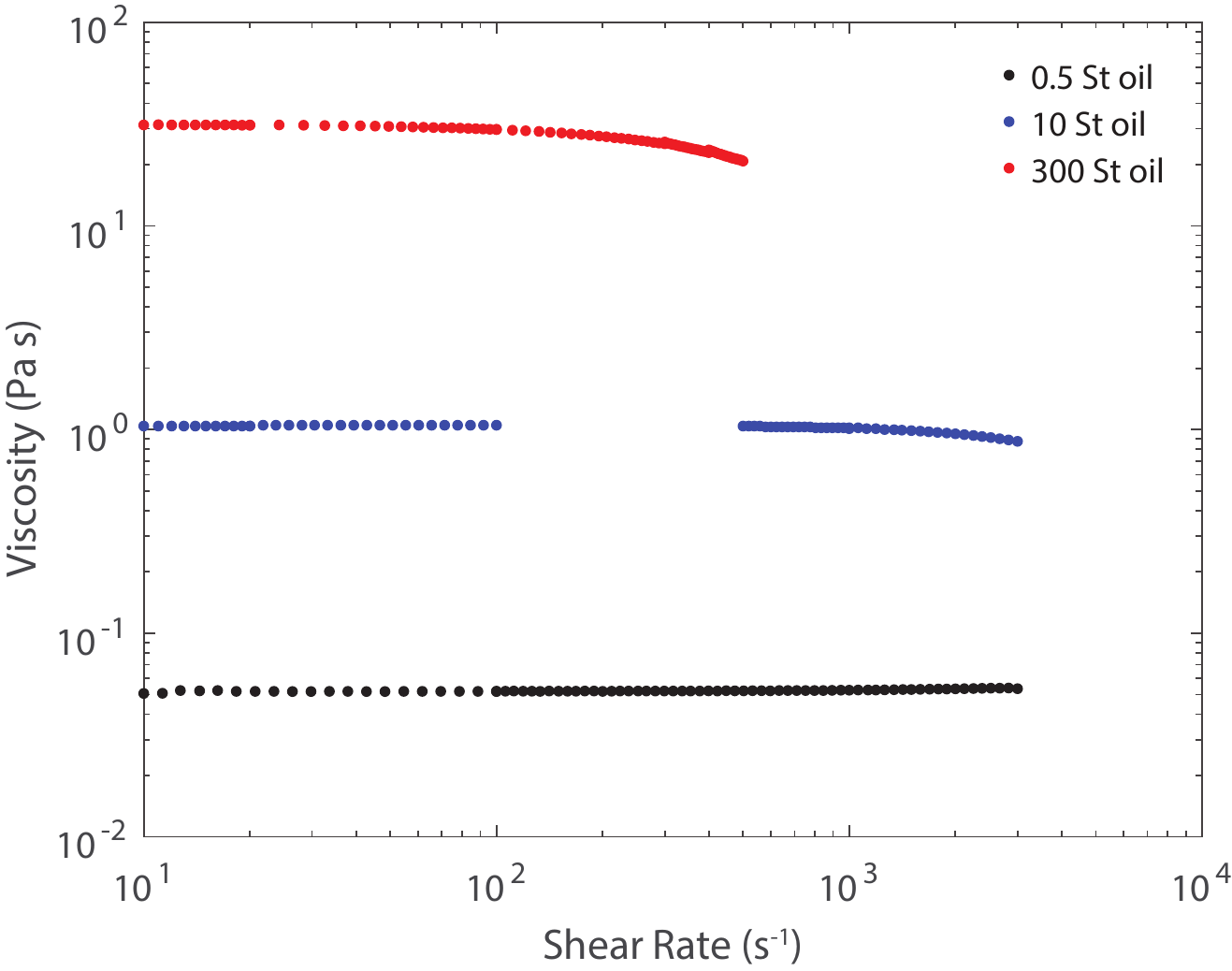}
\caption{Steady-state measurements of viscosity as a function of shear rate for the silicone oils used in this study.}
\label{fig:steadystateviscositytest}
\end{figure}

A series of steady-state rheometry measurements were taken in order to verify the viscosity of the silicone oils presented in Tab.~\ref{table1} and determine their shear rate sensitivity. The measurements were taken using an Anton Paar MCR 502 rotational rheometer with a concentric cylinder arrangement, which was able to achieve shear rates of \SI{3e3}{\per\second} for the low and intermediate-viscosity oils and \SI{5e2}{\per\second} for the high-viscosity oil. Figure~\ref{fig:steadystateviscositytest} shows the measured dynamic viscosity of the silicone oils as a function of the shear rate. The zero-shear-rate viscosity of the fluids agreed with the manufacturer provided values presented in Tab.~\ref{table1}. All three oils had a constant viscosity at low shear rates, which was seen to transition to a shear thinning behavior at higher shear rates for the intermediate and high-viscosity oils. The measured zero-shear-rate viscosity ($\eta_0$) and critical shear rate ($\dot{\gamma}_c$) for the onset of shear thinning are presented in Tab.~\ref{tab:table2}.

\begin{table}
\caption{Steady-state measurements of silicone oil viscosity as a function of strain rate.}
\label{tab:table2}
\begin{ruledtabular}
\begin{tabular}{lcc}
& $\eta_0$ & $\dot{\gamma}_c$\\
&(\SI{}{\pascal\second}) & (\SI{}{\per\second})\\
\hline\noalign{\smallskip}
0.5 St oil  & \num{5.17e-2} & \num{2e2} \\
10 St oil   & \num{1.05e0}  & \num{2e3} \\
300 St oil & \num{3.08e1}  & - \\
\end{tabular}
\end{ruledtabular}
\end{table}

The shear thinning behavior discussed above has been observed previously in PDMS oils.~\cite{Barlow1964,Ghannam1998,Carre2006,Vazquez2017} These oils are known to behave as Newtonian fluids at low shear rates, then begin to show shear thinning behavior at a critical shear rate which increases with decreasing zero-shear-rate viscosity.~\cite{Ghannam1998,Carre2006,Vazquez2017} The onset of shear thinning indicates that at these deformation rates, the viscoelastic properties of the PDMS oils, which result from interactions of the polymer molecules, begin to affect the flow. The high strain rates encountered in the experiments presented in this study ($\approx 10^4\, \mathrm{s}^{-1}$), will result in a very dynamic cavitation process with large bubble growth rates that are likely to induce viscoelastic material behavior.

\subsection{Plate-impact experiments}
\label{PIexperiments}
\begin{figure*}
\includegraphics[width=1.6\columnwidth]{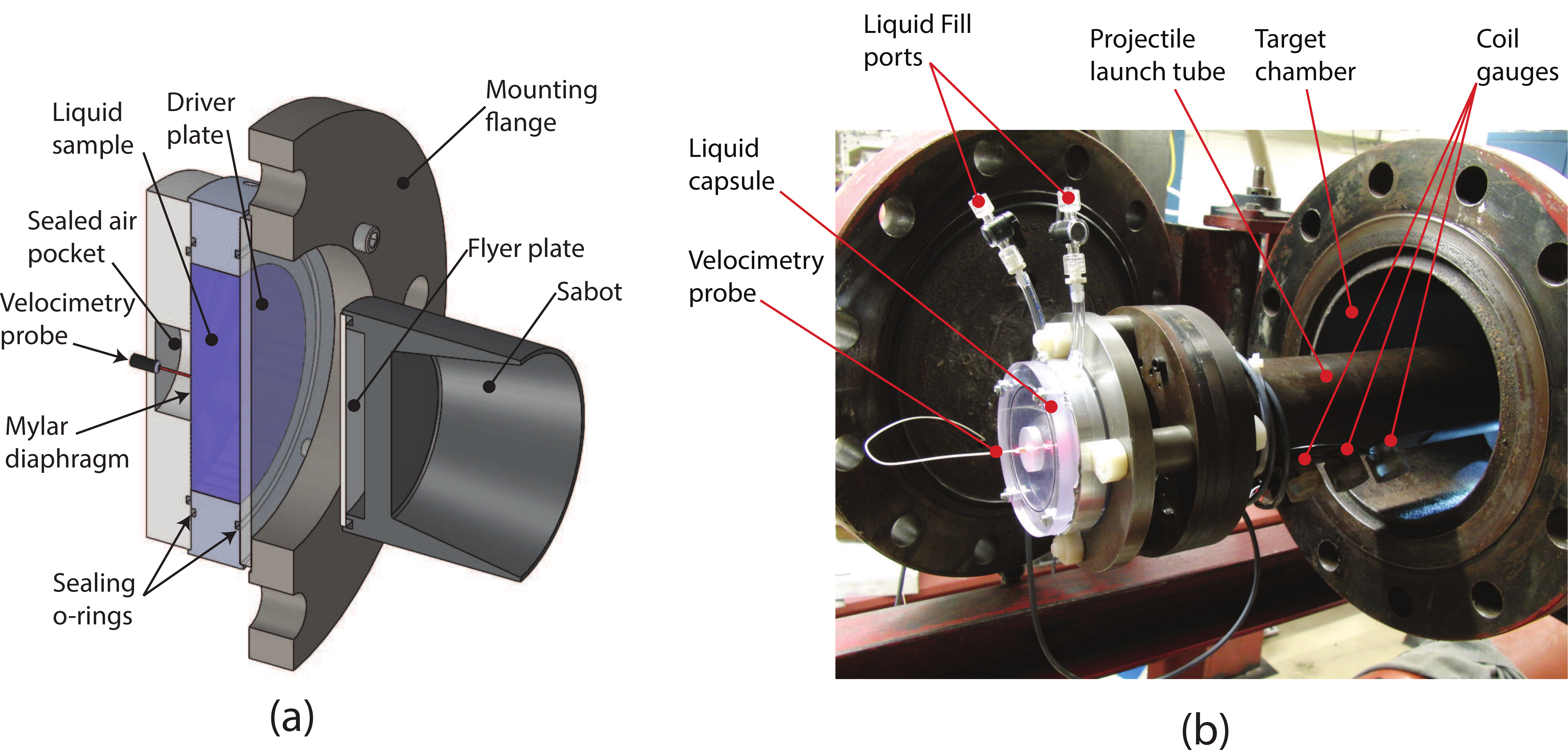}
\caption{(a) Schematic of the liquid capsule assembly, and (b) labelled picture of the assembly mounted onto the launch tube of the gas gun.}
\label{fig:targetassembly}
\end{figure*}

In this study, the planar impact of flyer plates onto liquid samples was used to determine the spall threshold of the PDMS oils. Poly(methyl methacrylate) (PMMA) or aluminum (Al~6061) flyer plates, held by a sabot, were launched from a 64~mm-bore single-stage gas gun, and impacted the liquid target assembly. A magnet was embedded in the sabot in order to monitor the impact velocity of the flyer plate using a series of magnetic coil gauges located at the end of the projectile launch tube. The front face of the target assembly consisted of a 3-mm-thick Polycarbonate driver plate which transmitted the shock wave into the test liquid upon impact. The liquid was contained within an aluminum ring which had fill ports through which the liquid was injected via Luer-Lok fittings. Prior to being injected in the capsule, the oils were degassed under vacuum (approximately 100~Pa) until free bubbles no longer appeared. The rear surface of the liquid was held in place by a 51-$\upmu$m-thick aluminized polyester sheet (Mylar), which formed a free surface through which the incident shock wave reflected to place the sample in tension. Although there was a slight impedance mismatch between the liquid sample and the Mylar, the short acoustic travel time across the film (on the order of \SI{30}{\nano\second}) ensured the film had no significant effect on the dynamics of the experiment. The rearmost portion of the target assembly held the collimating optical probe used to track the velocity of the Mylar surface. A sealed air gap between the Mylar surface and the probe allowed the liquid and Mylar to move freely during the experiment, which lasts on the order of \SI{1}{\micro\second}. The dimensions of the target assembly were chosen such that waves caused by the interaction of the impact-driven shock wave with the boundaries of the liquid did not reach the central axis in time to interfere with the spall measurement, thus ensuring that the strain remained uniaxial. The target assembly was fastened in a manner which ensured that the Mylar sheet was taut and free of wrinkles and was mounted onto a flange on the end of the gas gun launch tube in order to ensure a planar impact. Face-seal o-rings and stopcock valves were used to seal the target assembly, which allowed the target chamber to be evacuated to a 200~Pa Helium atmosphere before the shot, while maintaining a 1~atm environment within the target assembly. A labelled schematic of the projectile and target assembly is shown in Fig.~\ref{fig:targetassembly}(a), while a labeled picture of the target assembly mounted onto the gas gun can be seen in Fig.~\ref{fig:targetassembly}(b).

\begin{figure}
\includegraphics[width=0.9\columnwidth]{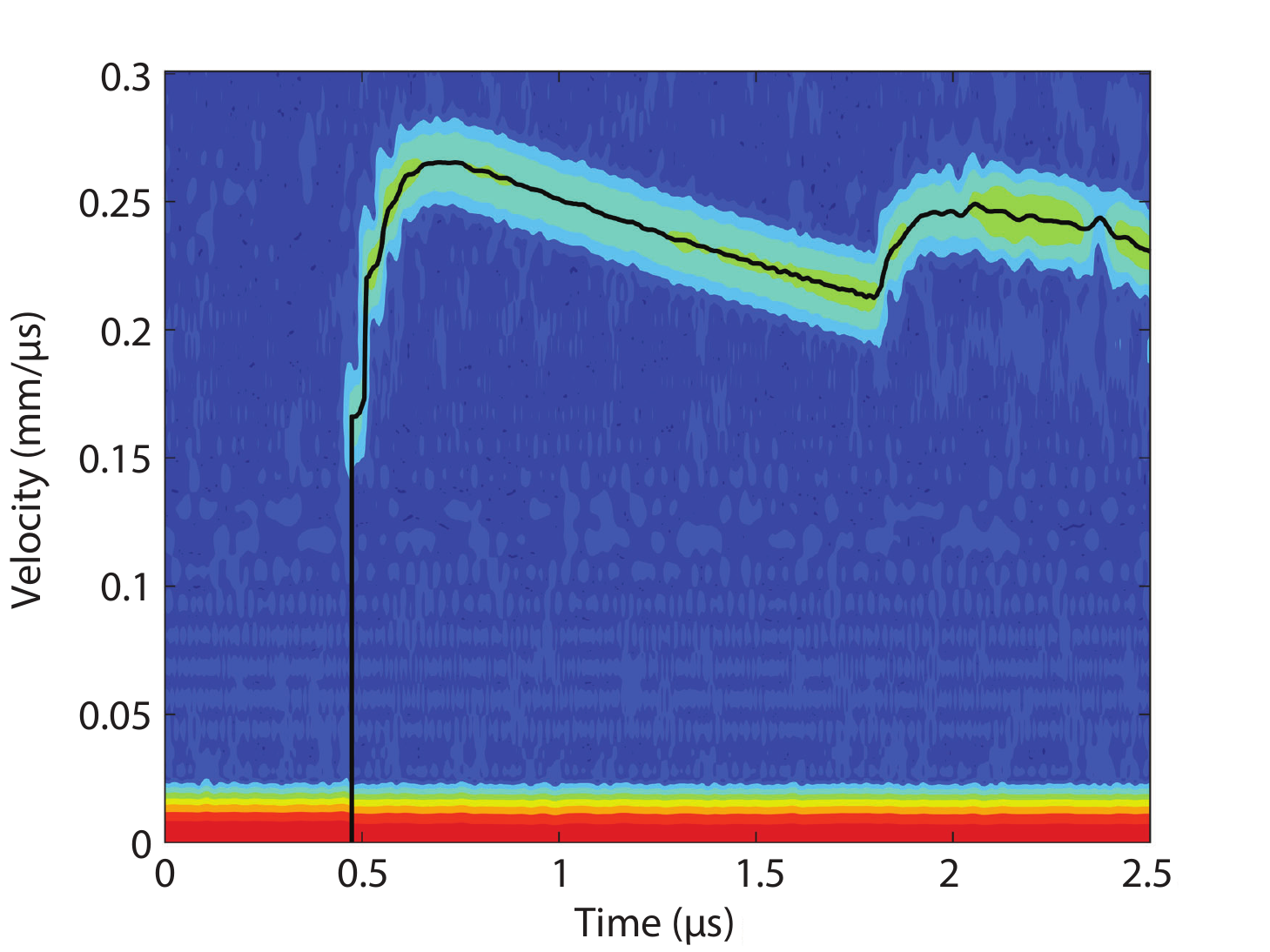}
\caption{Typical spectrogram obtained from performing a sliding window FFT on the PDV data. The black curve is the evolution in the free-surface velocity as a function of time, which is obtained from the analysis.}
\label{fig:samplespectrogram}
\end{figure}

Plate-impact experiments rely on measuring the time evolution of the free-surface velocity to determine the spall strength of the material being studied.~\cite{Antoun2003} The use of aluminized Mylar, which is reflective to infrared radiation, allowed the free-surface velocity to be monitored by a photonic Doppler velocimetry (PDV) system.~\cite{Strand2006} PDV measures surface velocity by observing the variation in beat frequency between a laser beam reflected off a moving surface and a reference beam. The system has a trade-off between time resolution and velocity accuracy, due to the fact that the beat frequency is obtained by performing a sliding window fast Fourier transform (FFT).~\cite{Dolan2010} A window sample size of 2000 to 3000 samples was used during the sliding FFT analysis, resulting in an estimated velocity uncertainty of 2~m/s and a time resolution of approximately 10~ns. Figure~\ref{fig:samplespectrogram}, shows a typical spectrogram produced by performing a sliding window FFT on the PDV data, where the high intensity contours represent the velocity of the free surface. The extracted velocity--time curve for the experiment is also shown in black. It should be noted that previously reported experiments~\cite{Huneault2018} related to this study used significantly thinner aluminized Mylar diaphragms (\SI{4}{\micro\metre}). The \SI{4}{\micro\metre} diaphragms had a lower reflectivity and a tendency to wrinkle which affected the PDV return signal, sometimes leading to noisy data or lost experiments. Increasing the Mylar thickness to \SI{51}{\micro\metre} improved the quality and consistency of the velocimetry data, and was not seen to affect the spall strength measurements.

\subsection{Analysis methods}
\label{analysis}
\begin{figure*}
\includegraphics[width=1.6\columnwidth]{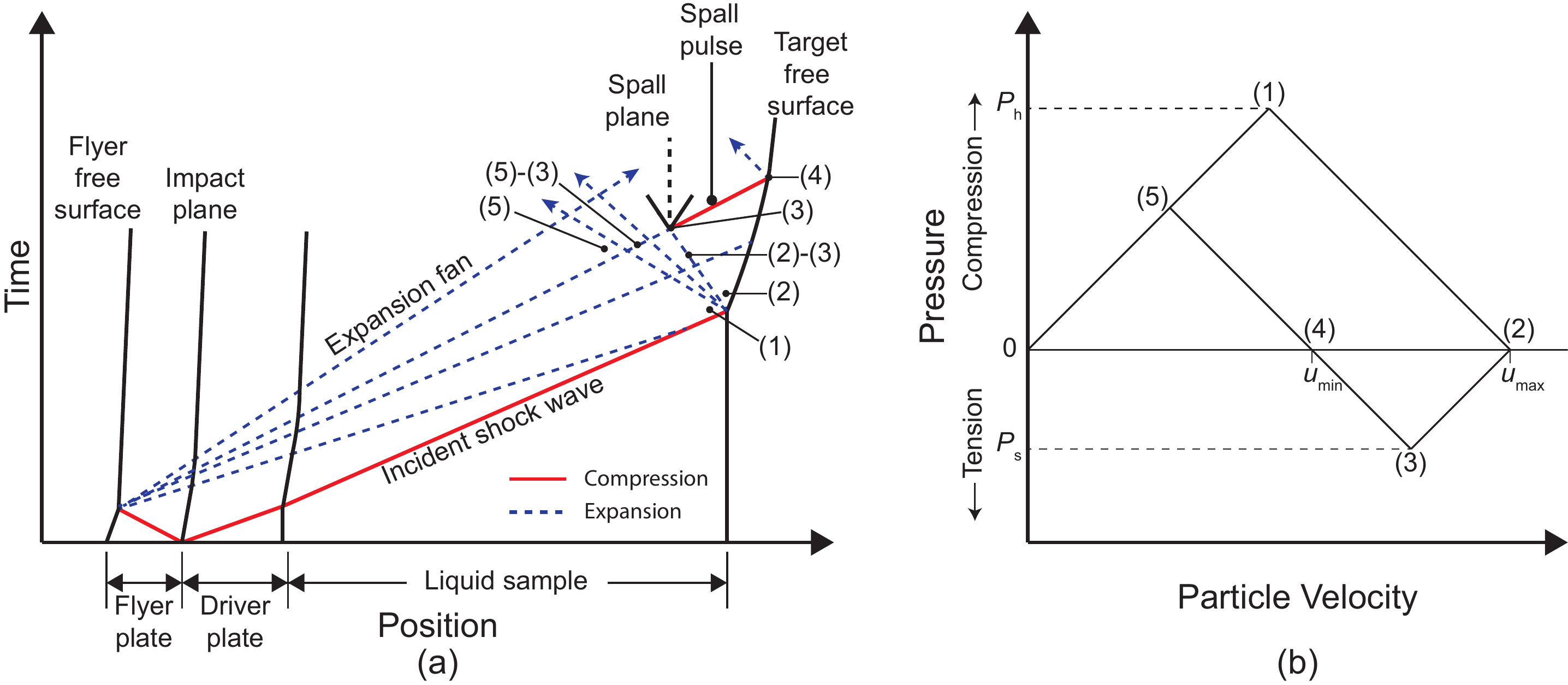}
\caption{(a) Position-time wave diagram of a plate-impact experiment which demonstrates the mechanism for shock induced cavitation. (b) Schematic of the evolution in the particle velocity and pressure within the silicone oil target during a plate-impact experiment. The key points in the loading path are labelled in part (a) of the figure.}
\label{fig:wavediagram}
\end{figure*}

This section will outline the methods used to analyze the free-surface velocity of the target liquid in order to determine the maximum pressure in the liquid prior to unloading, the rate at which tension increases in the liquid prior to cavitation, and the maximum tension in the liquid (i.e., spall strength). A schematic of the wave diagram for the plate-impact experiment is shown in Fig.~\ref{fig:wavediagram}(a). Initially, the impact drives a shock wave into both the driver plate and the flyer plate. The shock in the driver is transmitted into the liquid sample, while the shock in the relatively thin flyer plate rapidly reflects from its free surface as an expansion front, which is then transmitted into the driver and liquid. Nonlinearity allows the head of the expansion front from the flyer free surface to overtake the shock wave for sufficiently long travel distances. The thickness of the flyer plate and target capsule in these experiments have been chosen such that the shock is unsupported by the time it reaches the free surface (i.e., the shock is directly followed by an expansion front). The shock in the liquid reflects from the rear free surface as an expansion front. The two expansion fronts then interact with each other and generate a rapidly increasing tension in the liquid, which is eventually relieved by cavitation at a critical tension which corresponds to the spall strength of the liquid. The resulting increase in pressure caused by cavitation forms a compressive wave, known as a \textit{spall pulse}, which travels towards the rear free surface. 

\begin{figure}
\includegraphics[width=0.9\columnwidth]{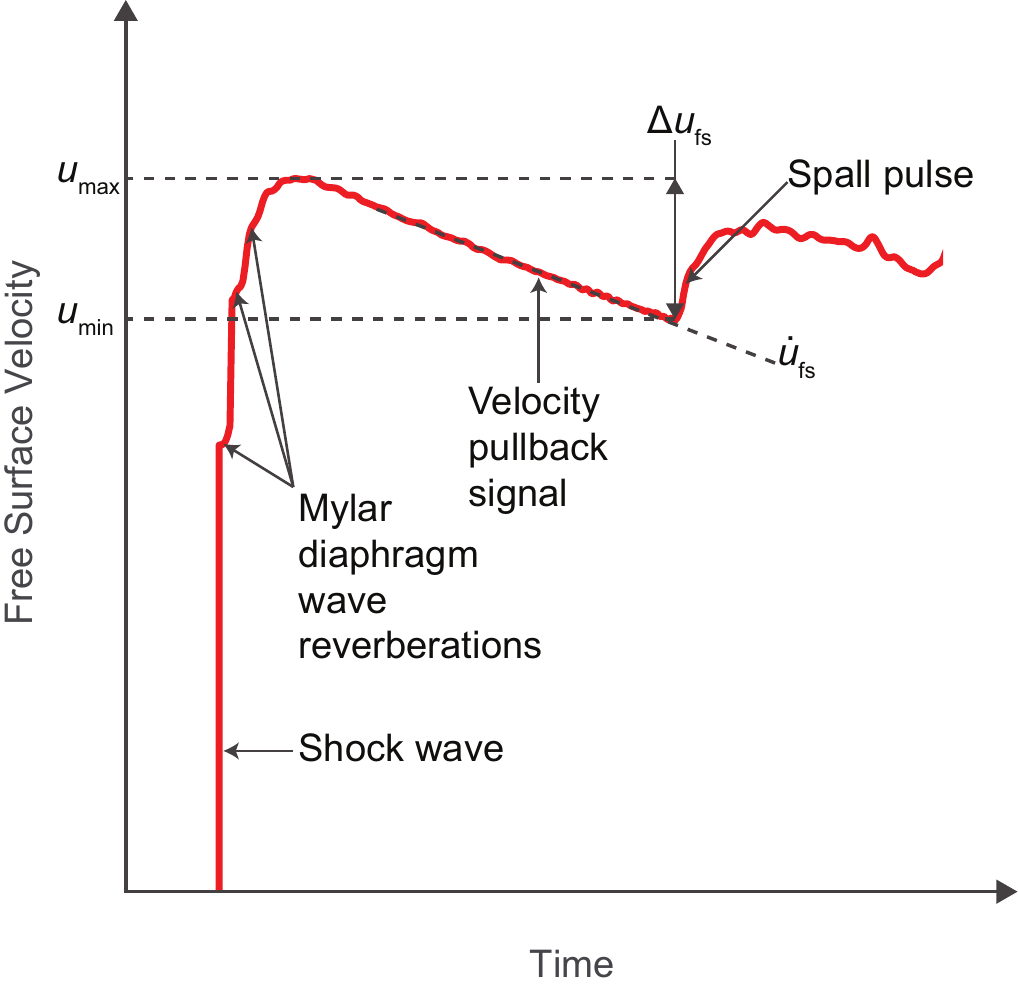}
\caption{Labeled schematic of the evolution in the silicone oil target free-surface velocity as a function of time for a plate-impact experiment.}
\label{fig:freesurfacevelocityschematic}
\end{figure}

The process described above is illustrated in the pressure--particle velocity plot in Fig.~\ref{fig:wavediagram}(b), while a representative example of the time evolution of the liquid free-surface velocity is shown in Fig.~\ref{fig:freesurfacevelocityschematic}, where the main features of the signal have been labeled. The arrival of the shock wave increases the pressure and particle velocity in the liquid to state~(1), which transitions to state~(2) as the reflected expansion front relieves the pressure and doubles the particle velocity. On the free-surface velocity trace, the shock wave manifests itself as a sharp increase in pressure, with the peak velocity ($u_\mathrm{max}$) corresponding to state~(2). As can be seen in Fig.~\ref{fig:freesurfacevelocityschematic}, wave reverberations in the Mylar sheet cause a slight delay in reaching the peak velocity behind the shock. The interaction of both expansion fronts further reduces the pressure into the tensile region (state~(2)-(3)), which manifests itself on the free-surface velocity trace as a nearly linear decrease in velocity. This velocity pullback signal has a slope which is proportional to the rate at which tension develops in the sample. The state of maximum tension ($P_\mathrm{s}$, state~(3)), corresponds to the point where tension is relieved by cavitation in the liquid, while state~(4) corresponds to the minimum free-surface velocity ($u_\mathrm{min}$) observed in Fig.~\ref{fig:freesurfacevelocityschematic}, which occurs at the arrival of the cavitation-induced spall pulse. 

From Fig.~\ref{fig:wavediagram}(a), it can be seen that cavitation is expected to initiate at the intersection of the tail of the rear free-surface expansion front (line~(2)-(3)) with the forward-moving characteristic where the critical tension is first reached (line~(3)-(4)-(5)). Therefore, in the acoustic approximation, the spall strength can be evaluated from the velocity pullback ($\Delta u_\mathrm{fs}=u_\mathrm{max}- u_\mathrm{min}$) recorded on the free surface-velocity trace using the following expression~\cite{Antoun2003}

\begin{equation}
\label{eq:1}
P_\mathrm{s} =\frac{1}{2}\rho_0c_0\Delta u_\mathrm{fs},
\end{equation}

\noindent where $\rho_0$ and $c_0$ are the ambient density and speed of sound of the liquid. Similarly, the strain rate is inferred by fitting a linear slope to the free-surface velocity pullback signal ($\dot{u}_\mathrm{fs}$)~\cite{Sheng2009}

\begin{equation}
\label{eq:2}
\dot{\epsilon}=\frac{-\dot{u}_\mathrm{fs}}{2c_0}.
\end{equation}

\noindent It is also of interest to estimate the maximum pressure experienced by the liquid target as it is shock compressed prior to unloading. The particle velocity behind the shock wave ($u_\mathrm{p}$) can be estimated from the maximum free-surface velocity during shock-up using the following relation~\cite{Walsh1955}

\begin{equation}
\label{eq:3}
u_ {\mathrm{p}}\approx \frac{1}{2}u_ {\mathrm{max}}.
\end{equation}

\noindent From the conservation of mass and momentum across a shock wave, the post-shock pressure ($P_\mathrm{h}$) can be expressed as

\begin{equation}
\label{eq:4}
P_ {\mathrm{h}}=\rho_ {\mathrm{0}}U_ {\mathrm{s}}u_ {\mathrm{p}},
\end{equation}

\noindent where $U_\mathrm{s}$ is the shock velocity. It is well known that most materials exhibit a linear relationship between shock velocity and particle velocity, known as the $U_ \mathrm{s}-u_ \mathrm{p}$ Hugoniot.~\cite{Cooper2002} Although this empirical relationship has not been determined for PDMS oils, it is possible to estimate the shock velocity from the ambient speed of sound of the liquid using the so-called \textit{universal Hugoniot} relationship proposed by Woolfolk et al.~\cite{Woolfolk1973}

\begin{equation}
\label{eq:5}
U_ {\mathrm{s}}=1.37c_0-0.37c_0\mathrm{exp}\left(\frac{-2u_\mathrm{p}}{c_0}\right)+1.62u_ \mathrm{p}.
\end{equation}

\noindent It has been shown that this relationship agrees well with a wide variety of liquids as long as the post-shock pressure remains below the bulk modulus of the liquid~\cite{Garrett2006}, which is the case in this study.

\section{RESULTS}
\label{results}
\begin{figure*}
\includegraphics[width=1.6\columnwidth]{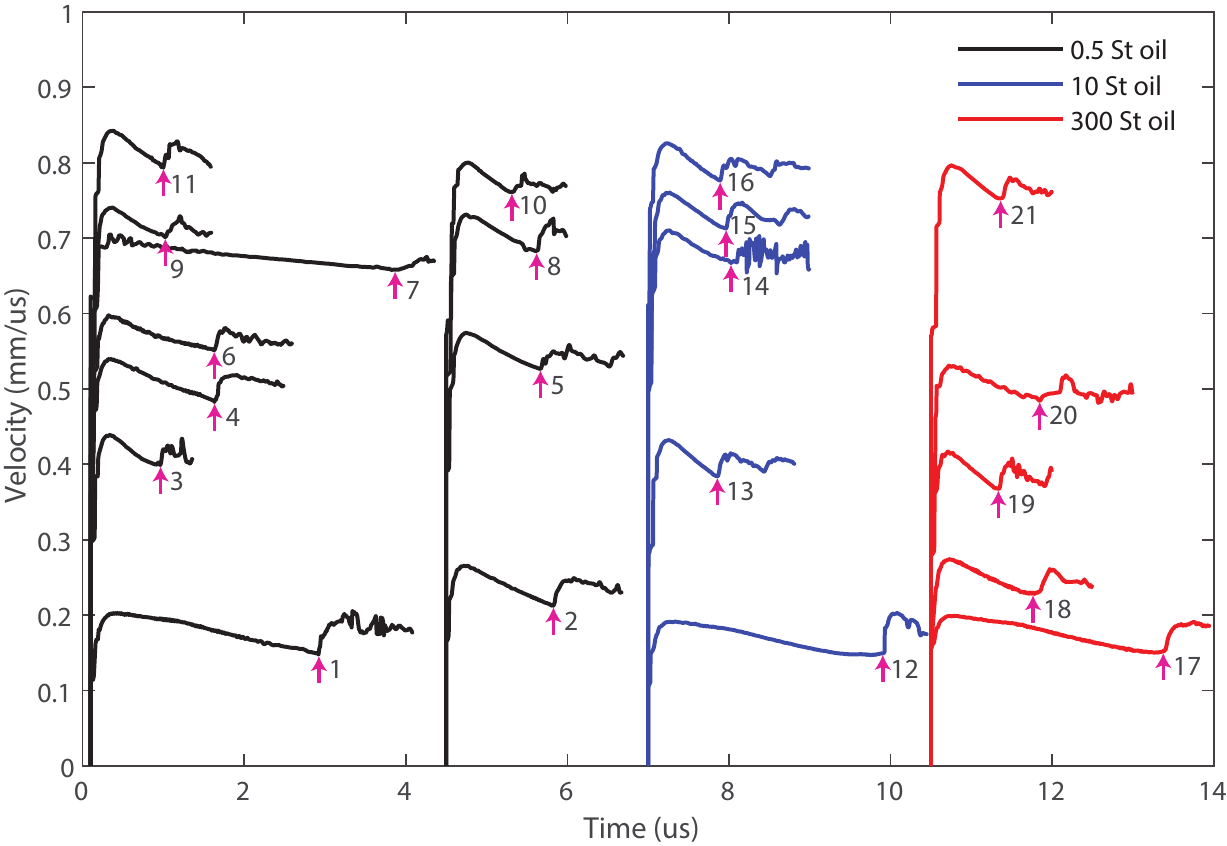}
\caption{Time-velocity plot of the free surface velocities for the plate-impact experiments. The time axis has been arbitrarily shifted to fit all the data on the same plot. Each curve is labeled by its shot number and the arrival of the spall pulse is marked by an arrow.}
\label{fig:freesurfacevelocities}
\end{figure*}

The test parameters, including the target liquid thickness ($w_\mathrm{l}$), the flyer plate thickness ($w_\mathrm{f}$), and the flyer plate impact velocity ($v_\mathrm{i}$), as well as the resulting maximum particle velocity ($u_\mathrm{max}$) at the free surface, calculated post-shock pressure ($P_\mathrm{h}$), pullback velocity ($\Delta u_\mathrm{fs}$), strain rate ($\dot{\epsilon}$), and measured spall strength ($P_\mathrm{s}$) for 21 plate-impact experiments performed on the PDMS oils are presented in Tab.~\ref{tab:table3}. The free-surface velocity profiles for the experiments are presented in Fig.~\ref {fig:freesurfacevelocities}, where the curves have been arbitrarily shifted on the time axis in order to display them on the same plot. A spall event, identified by an arrow in Fig.~\ref {fig:freesurfacevelocities}, was recorded for each test. As expected, an increase in the impact velocity resulted in greater shock pressures and peak free surface velocities. Similarly, experiments with aluminum flyers generated greater shock pressures than those with PMMA flyers at equivalent impact velocities due to the difference in the shock impedances of the materials. In general, greater impact velocities also led to greater strain rates.

\begin{table*}
\caption{Summary of plate-impact experiments.}
\label{tab:table3}
\begin{ruledtabular}
\begin{tabular}{lcccccccccc}
Shot & Target & $w_{\mathrm{l}}$ & Flyer & $w_\mathrm{f}$ & $v_\mathrm{i}$ & $u_\mathrm{max}$ & $P_\mathrm{h}$ & $\Delta u_\mathrm{fs}$ & $\dot{\epsilon}$ & $P_\mathrm{s}$\\
No. & Liquid & (mm) & Material & (mm) & (\SI{}{\metre\per\second}) & (\SI{}{\metre\per\second}) & (\SI{}{\mega\pascal}) & (\SI{}{\metre\per\second}) & ($10^4$\SI{}{\per\second}) & (\SI{}{\mega\pascal}) \\
\hline\noalign{\smallskip}
1 & 0.5 St    & 13 & Al 6061   & 4.8 & 124 & 203 & 120 & 54 & 1.1 & 26 \\ 
2 & 0.5 St    & 14 & PMMA     & 1.6 & 445 & 265 & 166 & 53 & 2.4 &  25\\ 
3 & 0.5 St    & 9   & PMMA     & 1.6 & 628 & 439 & 314 & 39 & 3.7 &  19\\ 
4 & 0.5 St    & 17 & Al 6061   & 6.3 & 395 & 540 & 413 & 57 & 2.4 &  27\\ 
5 & 0.5 St    & 13 & Al 6061   & 4.8 & 402 & 574 & 450 & 48 & 3.0 &  23\\ 
6 & 0.5 St    & 9   & Al 6061   & 4.1 & 387 & 598 & 475 & 46 & 2.3 &  22\\ 
7 & 0.5 St    & 17 & Al 6061   & 6.3 & 592 & 705 & 596 & 47 & 0.6 &  23\\ 
8 & 0.5 St    & 17 & Al 6061   & 6.3 & 581\footnotemark[1] & 731 & 627 & 48 & 3.2 &  23\\ 
9 & 0.5 St    & 14 & Al 6061   & 4.6 & 545 \footnotemark[1]& 741 & 639 & 39 & 3.2 &  19\\ 
10 & 0.5 St  & 9   &  Al 6061 & 4.1 & 550 & 800 & 713 & 39 & 4.1 &  19\\ 
11 & 0.5 St  & 9   & Al 6061 & 4.1 & 573 & 842 & 767 & 49 & 4.1 &  24\\ 
12 & 10 St   & 13 & Al 6061 & 4.8 & 120 & 191 & 114 & 44 & 1.1 &  22\\ 
13 & 10 St   & 9   & PMMA   & 1.6 & 622 & 432 & 311 & 48 & 4.4 &  23\\ 
14 & 10 St   & 14 & Al 6061 & 4.6 & 545 & 710 & 609 & 43 & 3.4 &  21\\ 
15 & 10 St   & 17 & Al 6061 & 6.3 & 570 & 760 & 671 & 48 & 3.8 &  23\\ 
16 & 10 St   & 9   & Al 6061 & 4.1 & 564 & 826 & 754 & 49 & 4.1 &  24\\ 
17 & 300 St & 13 & Al 6061 & 4.8 & 123 & 199 & 119 & 49 & 1.0 &  24\\ 
18 & 300 St & 14 & PMMA   & 1.6 & 442 & 274 & 176 & 45 & 2.6 &  22\\ 
19 & 300 St & 9   & PMMA   & 1.6 & 605 & 417 & 297 & 49 & 4.5 &  24\\ 
20 & 300 St & 17 & Al 6061 & 6.3 & 379 & 531 & 409 & 47 & 2.4 &  23\\ 
21 & 300 St & 13 & Al 6061 & 4.8 & 572 & 796 & 716 & 44 & 4.4 &  21\\ 
\end{tabular}
\end{ruledtabular}
\footnotetext[1]{Projectile velocity estimated from experiments with the same projectile and launcher parameters.}
\end{table*}

\begin{figure}
\includegraphics[width=0.9\columnwidth]{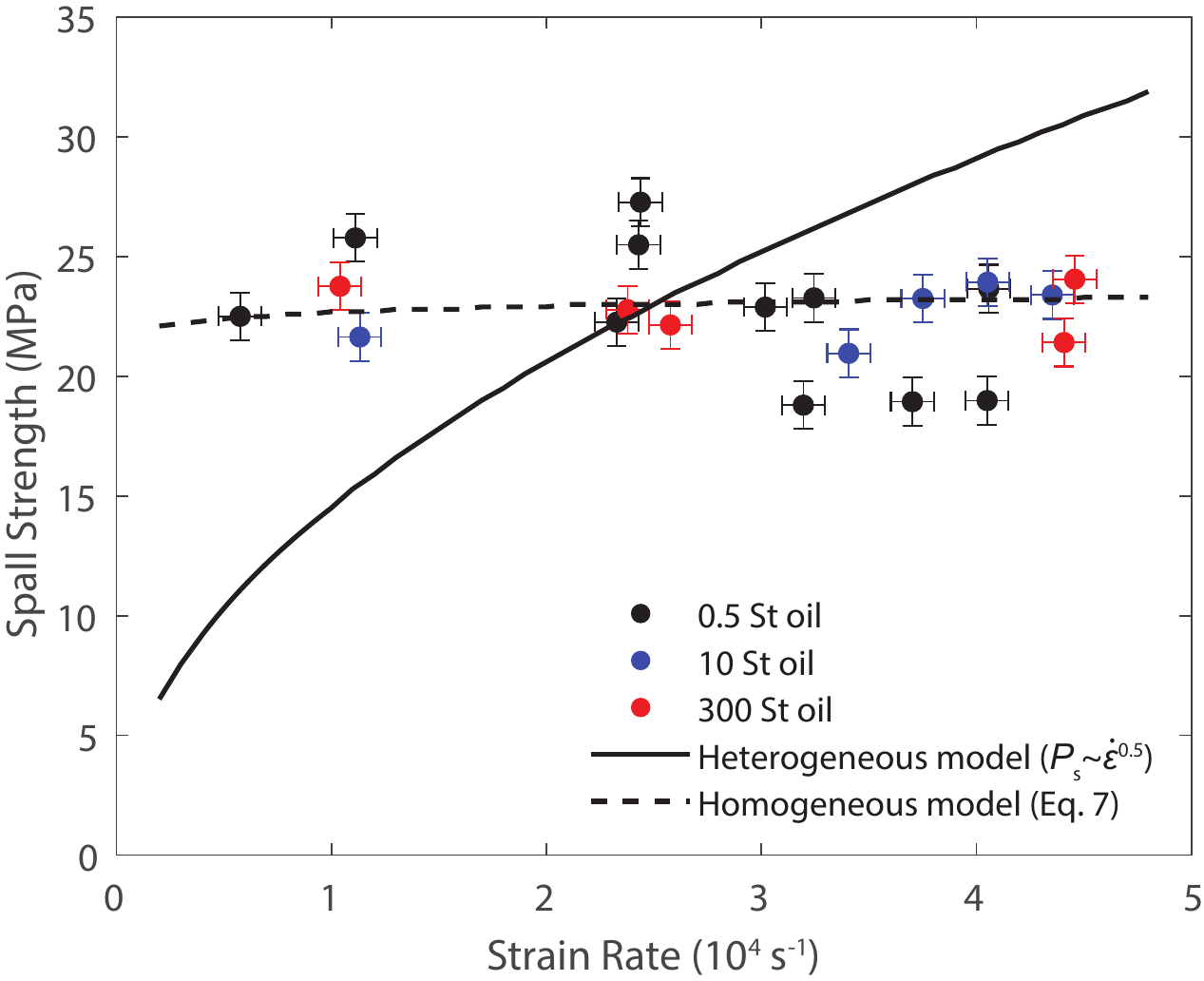}
\caption{Spall strength as a function of strain rate for the plate-impact experiments. Also shown is the expected sensitivity of spall strength to strain rate based on homogeneous and heterogeneous cavitation models.}
\label{fig:spallvsstrainrate}
\end{figure}

\begin{figure}
\includegraphics[width=0.9\columnwidth]{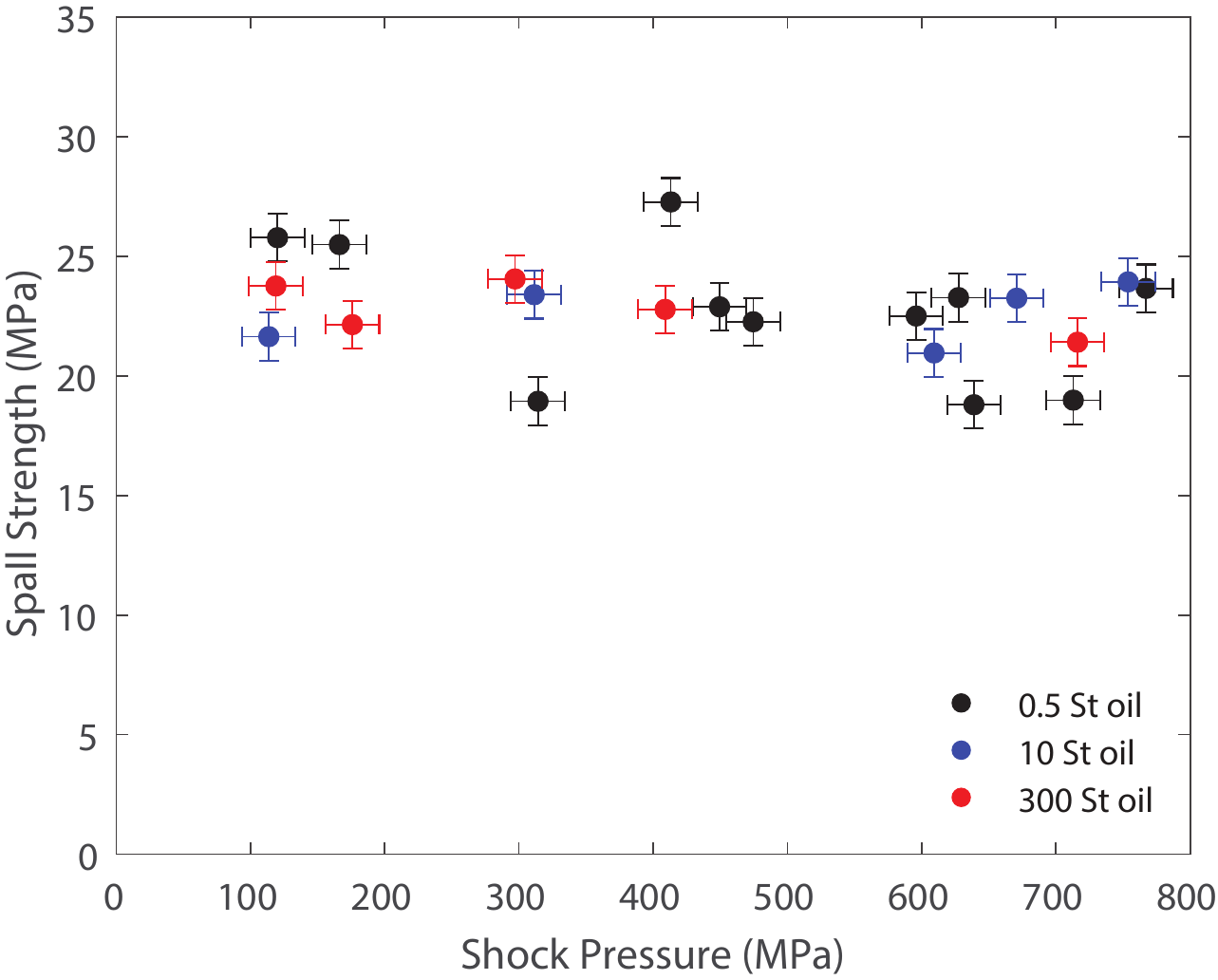}
\caption{Spall strength as a function of shock pressure for the plate-impact experiments.}
\label{fig:spallvspressure}
\end{figure}

\begin{figure}
\includegraphics[width=0.9\columnwidth]{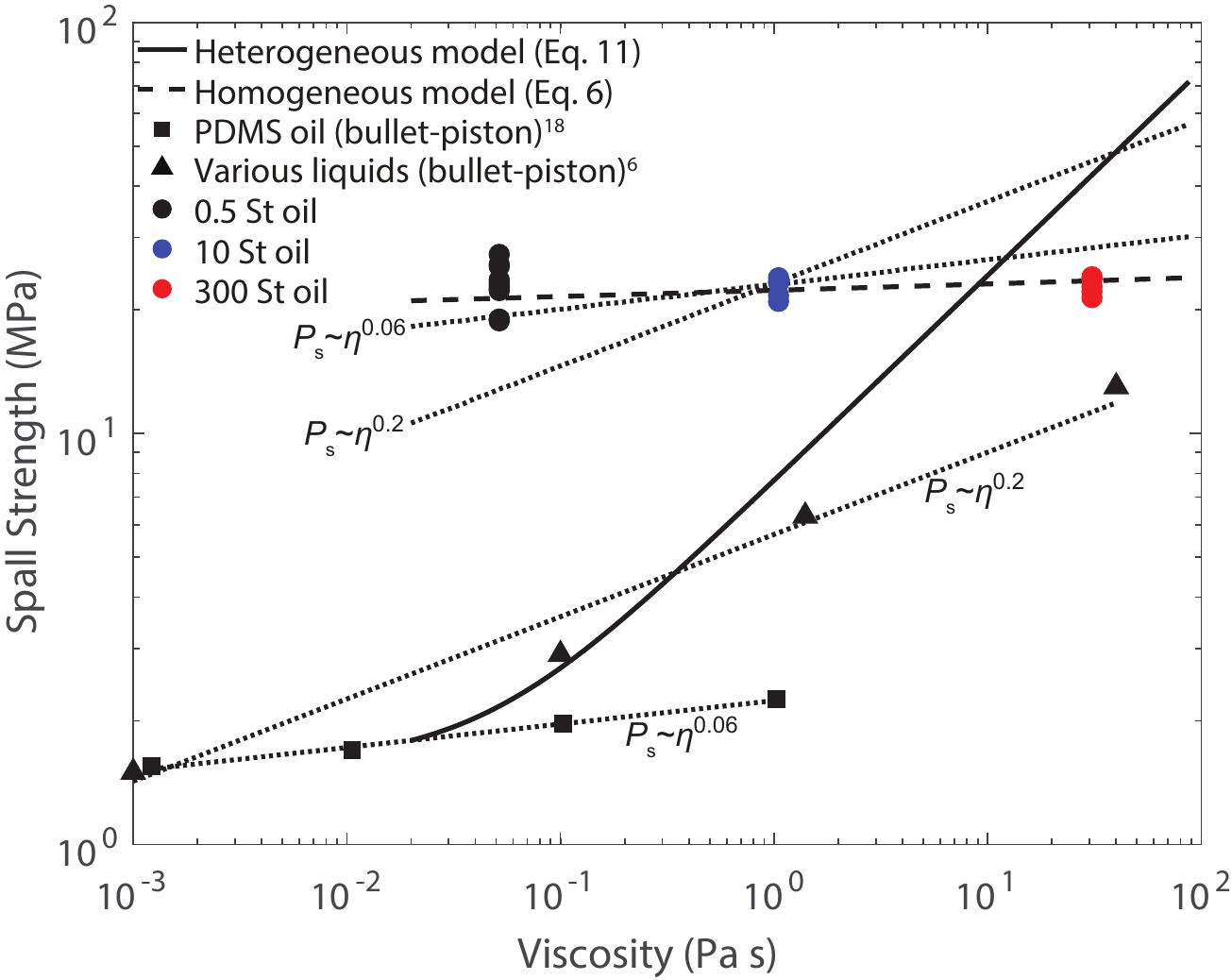}
\caption{Measured spall strength from the plate-impact experiments as a function of the zero-shear-rate viscosity ($\eta_0$) of the PDMS oils. Also shown is the expected sensitivity of spall strength to viscosity based on homogeneous and heterogeneous cavitation models,  and previously published data and viscosity correlations from Bull~\cite{Bull1956} and Couzens and Trevena.~\cite{Couzens1974}}
\label{fig:spallvsviscosity}
\end{figure}

The measured spall strength for the plate-impact experiments is plotted as a function of strain rate ($\dot{\epsilon}$) in Fig.~\ref{fig:spallvsstrainrate} for the three types of PDMS oils. Also included in Fig.~\ref{fig:spallvsstrainrate} are curves based on homogeneous and heterogeneous nucleation models described in Section~\ref {theory}. The experimental data does not show a strain rate dependency for either of the three silicone oils. The measured spall strength is also plotted as a function of the shock pressure ($P_\mathrm{h}$) in Fig.~\ref{fig:spallvspressure}. As can be seen, the measured spall strength is unaffected by the wide variation in shock pressure. In Fig.~\ref{fig:spallvsviscosity}, the measured spall strength is plotted as a function of the zero-shear-rate viscosity for the three PDMS oils. Curves based on homogeneous and heterogeneous cavitation models described in Section~\ref {theory} are also shown, as well as the data and correlations between critical tension and viscosity obtained from bullet-piston experiments of Bull~\cite{Bull1956} and Couzens and Trevena.~\cite{Couzens1974} A second set of power-law correlations based on the results of Bull~\cite{Bull1956} and Couzens and Trevena~\cite{Couzens1974} ($P_\mathrm{s}\sim \eta^{0.2}$ and $P_\mathrm{s}\sim \eta^{0.06}$, respectively) are also plotted such that they intercept the average spall strength value of \SI{23}{\mega\pascal} at \SI{9.71e-1}{\pascal\second}, which corresponds to the viscosity of the 10~St oil. As can be seen, the measured spall strength in this work showed no clear correlation with the viscosity of the oils. The average spall strength was 22.7~MPa for the 0.5~St silicone oil, 22.6~MPa for the 10~St oil, and 22.8~MPa for the 300~St oil.

\section{CAVITATION MODELS}
\label{theory}
The spall strength obtained from plate-impact experiments on liquid targets is a measure of the critical tension required to induce cavitation bubble growth. A liquid in tension is in a metastable state, where bubble nucleation and growth, the mechanism by which tension is relieved, is stabilized by the surface energy required to grow these voids. Bubbles in a liquid in tension ($P$) must be larger than a certain critical radius ($R_\mathrm{c}=2\sigma / P$), before it is energetically favorable for them to grow. The vapor pressure inside the bubble has been neglected in the critical radius expression above due to its small contribution compared to $P$, which is on the order of \SI{10}{\mega\pascal} in this work. $R_\mathrm{c}$ corresponds to the radius at which the work of formation of the bubble is maximum ($W_\mathrm{max}$). As was discussed in Section~\ref{introduction}, the cavitation bubbles which grow to relieve tension in the liquid may originate from pre-existing flaws, such as free bubbles or bubbles attached to contaminant particles, but bubbles may also nucleate from random molecular fluctuations in a homogeneous cavitation process. A further consideration in spall experiments is that tension is applied on a timescale which can be comparable to the rate of cavitation bubble growth. This can allow the liquid to remain in a state of tension despite the growth of bubbles if their density or growth rate is insufficient to relieve the tension that is accumulating in the sample.~\cite{Utkin1997,Utkin2006} The rate of void growth in a spalling material is typically characterized by the damage rate ($\dot{V}_\mathrm{v}$), which is a measure of the time rate of change of the specific volume of voids in the target material. For plate-impact experiments, it has been shown that the damage rate must exceed the strain rate by a factor of four or greater ($\dot{V}_\mathrm{v}>4\dot{\epsilon}/\rho_0$) in order to relieve tension in the sample and form the spall pulse on the free-surface velocity signal.~\cite{Antoun2003} The spall strength of a liquid may therefore be affected by its physical properties, the distribution of pre-existing flaws, and the rate of loading. This section will present models which can be used to predict the spall strength and its dependence on strain rate and viscosity for both homogeneous and heterogeneous cavitation nucleation.

\subsection{Homogeneous nucleation}
\label{homogeneous}
In a homogenous nucleation process, random molecular fluctuations allow bubbles to overcome the energy barrier created by surface tension in order to reach the critical radius and continue to grow. According to the theory of nucleation, the rate at which bubbles of a critical radius form is proportional to $\mathrm{exp}(-W_\mathrm{max}/kT)$, where $k$ is the Boltzmann constant, and $T$ is the temperature.~\cite{Fisher1948,Zeldovich1992,Carlson1975} The work of formation needed to reach the critical radius therefore provides a barrier to cavitation and stabilizes the liquid phase. Following the approach of Fisher~\cite{Fisher1948} and Zeldovich~\cite{Zeldovich1992}, an expression can be obtained which expresses the volumetric rate of formation of critical bubbles as a function of the tension in the fluid~\cite{Bogach2000},

\begin{equation}
\label{eq:6}
J=\frac{N_0\sigma}{\eta}\left(\frac{\sigma}{kT}\right)^{\frac{1}{2}}\mathrm{exp}\left(\frac{-16\pi \sigma^3}{3P^2kT}\right),
\end{equation}

\noindent where $N_0$ is the number of molecules per unit volume. As noted by Fisher and Zeldovich, Eq.~\ref{eq:6} can be used to determine the critical tension ($P=P_\mathrm{s}$) for cavitation by substituting a reasonable value for the nucleation rate ($J$). Indeed, Fisher~\cite{Fisher1948} showed that varying the nucleation rate over a range of \num{e-15} to \num{e18} only changes the value of critical tension obtained in the calculation by a factor of 1.58. The exponential factor is the dominant term in Eq.~\ref{eq:6}, and it should therefore be expected that the critical tension is relatively insensitive to the viscosity of the fluid or the rate at which tension is applied. Bogach and Utkin~\cite{Bogach2000} noted that from Eq.~\ref{eq:6}, the relationship between the critical tension and strain rate should follow the empirical expression below

\begin{equation}
\label{eq:7}
P_\mathrm{s}\approx \frac{A}{\left[\mathrm{ln}\left(\frac{B}{\dot{\epsilon}}\right)\right]^{\frac{1}{2}}},
\end{equation}

\noindent where $A$ and $B$ are constants. Model curves based on the homogeneous nucleation theory presented above have been included in Fig.~\ref{fig:spallvsstrainrate} (Eq.~\ref{eq:7}, $A$~=~\num{1.24e8}, $B$~=~\num{1.00e17}) and Fig.~\ref{fig:spallvsviscosity} (Eq.~\ref{eq:6}, $J$~=~\SI{1}{\per\metre\cubed\second}). As can be seen, according to the steady-state homogeneous nucleation theory presented above, the spall strength should be expected to marginally increase with increasing strain rate and viscosity.

\subsection{Heterogeneous nucleation}
\label{heterogeneous}
Any real fluid will contain a number of heterogeneities from which cavitation can nucleate.~\cite{Kedrinskii2005} A cavitation process dominated by the growth of pre-existing bubbles should be expected to be strongly influenced by the dynamics of the bubbles. The time dependent evolution in the radius ($R$) of a single bubble in an effectively infinite liquid is described by the Rayleigh-Plesset equation.~\cite{Plesset1977} Below is the Rayleigh-Plesset equation for a Newtonian fluid where the liquid is under a far-field tension ($P$) and the internal bubble pressure and mass diffusion across the bubble interface are negligible
 
\begin{equation}
\label{eq:8}
R\ddot{R}+\frac{3}{2}(\dot{R})^2+\frac{4\eta}{\rho_0R}\dot{R}+\frac{2\sigma}{\rho R}=\frac{P}{\rho_0}.
\end{equation}

\noindent It has been noted that for the case of a system subjected to a large sudden increase in tension, such as that in a spall experiment, the initial normalized growth rate of a bubble that is significantly larger than the critical radius can be expressed as the following~\cite{Bull1956,Erlich1971,Bogach2000}
 
\begin{equation}
\label{eq:9}
\frac{\dot{R}}{R}=\frac{P}{4\eta}.
\end{equation}

\noindent The volumetric growth rate of a single bubble ($\dot{V}_\mathrm{b}$), can then be expressed by the simple expression below~\cite{Bogach2000}

\begin{equation}
\label{eq:10}
\dot{V}_\mathrm{b}=\frac{\pi R^3P}{\eta}.
\end{equation}

\noindent From Eq.~\ref{eq:10}, it can be seen that heterogeneous cavitation bubble growth should be expected to strongly depend on the viscosity of the fluid. Obtaining an estimate of the spall strength from the approach described above poses a number of challenges, notably determining the distribution of pre-existing bubbles, and considering the effect of neighboring bubbles in a multi-bubble system. 

Grady~\cite{Grady1988} proposed a unique approach to estimating the spall strength of dynamically stretched materials, based upon the criteria that the onset of fracture should be expected to occur when it is energetically favorable for the material to break apart. More precisely, failure in a dynamically stretched liquid is assumed to occur when the stored energy (elastic and kinetic) in the stretching liquid is equal to the viscous dissipation and surface energy produced in the fracture (cavitation) process.~\cite{Grady1988} Inherently, this approach assumes that the material is favorably disposed to fracture, which for a liquid means that there are pre-existing bubbles from which cavitation can nucleate.~\cite{Grady1988} The following expression, which relates the spall strength to the strain rate for the experiment ($\dot{\epsilon}$) and the surface tension ($\sigma$), viscosity ($\eta$), and speed of sound ($c_0$) of the liquid, can be obtained from the energy based approach~\cite{Grady1988}

\begin{equation}
\label{eq:11}
P_\mathrm{s}^3-2\eta \rho c_0^2\dot{\epsilon} P_\mathrm{s}-6\rho^2c_0^3\sigma\dot{\epsilon}=0.
\end{equation}

\noindent The second and third term of Eq.~\ref{eq:11} are the viscous dissipation and surface energy contributions, respectively. As can be seen, the viscous dissipation term is more sensitive to strain rate that the surface energy term ($P_\mathrm{s}\sim \eta^{1/2}$ versus $P_\mathrm{s}\sim \eta^{1/3}$), which means that as the strain rate is increased, the spall strength transitions from being dominated by the contribution of surface energy to that of viscous dissipation. A transition strain rate ($\dot{\epsilon}_\mathrm{t}$), can be defined as the rate at which the viscous and surface energy are equal~\cite{Grady1988}

\begin{equation}
\label{eq:12}
\dot{\epsilon}_\mathrm{t}=\frac{9}{2}\frac{\rho\sigma^2}{\eta^3}.
\end{equation}

\noindent The transition strain rate corresponds to \num{1.7e4}, \num{2.2}, and \SI{7.9e-5}{\per\second} for the low, intermediate, and high-viscosity silicone oils, respectively. For the strain rates encountered in experiments ($\dot{\epsilon}\approx$\SI{e4}{\per\second}), the oils, particularly the intermediate and high-viscosity samples, can be considered to be in the viscous dissipation dominated regime, where the spall strength can be expressed as

\begin{equation}
\label{eq:13}
P_\mathrm{s}=(2\eta\rho c_0^2\dot{\epsilon})^{\frac{1}{2}}.
\end{equation}

\noindent In this regime, the spall strength is seen to have a strong dependence on both strain rate and viscosity ($P_\mathrm{s}\sim \dot{\epsilon}^{1/2}$, $P_\mathrm{s}\sim \eta^{1/2}$). The dependence of spall strength on strain rate based on the viscous limit of the Grady heterogeneous nucleation model is shown in Fig.~\ref{fig:spallvsstrainrate}, while the relationship between the spall strength and the viscosity of the silicone oils predicted by Eq.~\ref{eq:11} is shown in Fig.~\ref{fig:spallvsviscosity}. As can be seen, the heterogeneous nucleation theory presented above predicts a significant variation in spall strength with increasing strain rate and viscosity.

\section{DISCUSSION}
\label{discussion}
The spall strength values reported in Section~\ref{results} are comparable to measurements reported in previous plate-impact studies performed on similar liquids, where the critical tension was also found to be on the order of \SIrange{10}{100}{\mega\pascal}.~\cite{Erlich1971,Carlson1975,Bogach2000,Utkin2010} The spall strength of the three PDMS oils remained relatively constant over the range of strain rates and shock pressures probed in this work. As can be seen in Fig.~\ref{fig:spallvsstrainrate}, the low sensitivity to strain rate is consistent with the homogeneous cavitation model but does not agree with the heterogeneous cavitation model of Eq.~\ref{eq:11}. Utkin and Sosikov~\cite{Utkin2010} noted a similar lack of strain rate dependence in plate-impact experiments performed with a range of liquids including water, methanol, and hexane, which was attributed to homogeneous cavitation nucleation. The absence of a correlation between spall strength and shock pressure further suggests that pre-existing bubbles do not play an important role in determining the critical tension at which the oils cavitate for the conditions probed in this work. The significant variation in shock pressure would be expected to affect the size of the voids due to their collapse prior to tensile loading. The results also showed that the spall strength of the PDMS oils was unaffected by their zero-shear-rate viscosity, with all three oils having an equivalent average spall strength. A comparison of the plate-impact data with the model curves for $P_\mathrm{s}$ as a function of viscosity in Fig.~\ref{fig:spallvsviscosity}, shows that the insensitivity of the spall strength to viscosity is captured by the homogeneous cavitation model, whereas the heterogeneous cavitation model suggests that the variation in the viscosity of the silicone oils should result in an order of magnitude increase in spall strength. The agreement between the spall strength measurements and Eq.~\ref{eq:6}, which has no free parameters, again suggests that the dominant cavitation mechanism in these experiments is homogeneous nucleation of bubbles via molecular fluctuations.

The relationship between spall strength and viscosity observed in this work does not agree with previously published results from bullet-piston experiments~\cite{Bull1956,Couzens1974}, where a power-law correlation was seen between viscosity and critical tension. Indeed, Couzens and Trevena~\cite{Couzens1974} studied a similar system of silicone oils with nominal kinematic viscosities ranging from 0.01~St  to 10~St and observed a power law dependence between the critical tension and viscosity ($P_\mathrm{s}\sim\eta^{0.06}$). It is interesting to note that the critical tension measured during plate-impact experiments is typically an order of magnitude greater than values obtained using bullet-piston experiments. This can be seen in Tab.~\ref{tab:table4}, where the measured critical tensions obtained from bullet-piston and plate-impact experiments are compared for a number of liquids. This comparison indicates that the critical tension which induces cavitation increases significantly as the strain rate is increased from \SI{e2}{\per\second} in the bullet-piston studies~\cite{Grady1988} to \SI{e4}{\per\second} for plate-impact experiments. However, the results presented in this paper show that a sensitivity to loading rate is no longer observed at the strain rates encountered in plate-impact experiments. The lack of correlation between viscosity, strain rate, and spall strength observed in the results presented in this paper indicates a possible transition in the mechanism of cavitation nucleation as the rate at which tension is applied is increased from \SI{e2}{\per\second} (bullet-piston) to \SI{e4}{\per\second} (plate-impact).

\begin{table}
\caption{Comparison of critical tension measurements from bullet-piston and plate-impact experiments.}
\label{tab:table4}
\begin{ruledtabular}
\begin{tabular}{lcc}
Liquid & $P_\mathrm{s}$ (Bullet-Piston) & $P_\mathrm{s}$ (Plate-Impact)\\
& (\SI{}{\mega\pascal}) & (\SI{}{\mega\pascal})\\
\hline\noalign{\smallskip}
PDMS oil (10 St) & 2.3~\cite{Couzens1974} & 21-24$\footnotemark[1]$ \\
\noalign{\smallskip}
Water & 0.9-1.5~\cite{Couzens1969} & 32~\cite{Huneault2018} \\
&1.8~\cite{Bull1956b} &22.5-46.0~\cite{Bogach2000} \\
\noalign{\smallskip}
Glycerol & 6.3~\cite{Bull1956b} & 22-24~\cite{Erlich1971} \\
& & 57-142~\cite{Utkin2003} \\
\end{tabular}
\end{ruledtabular}
\footnotetext[1]{Data presented in this work.}
\end{table}

The mechanisms of spall failure in liquids bears resemblance to ductile spall failure in solids, which also proceeds via a mechanism of void nucleation, growth, and coalescence.~\cite{Grady1988} In metals, failure sites are known to nucleate at grain boundaries, inclusions, and second phase particles.~\cite{Curran1987,Antoun2003} Experiments performed on single crystal samples show a significant increase in spall strength~\cite{Razorenov2006,Razorenov2007} due to the fact that failure sites must now nucleate within the lattice at sub-microscopic heterogeneities such as dislocation pileups.~\cite{Antoun2003,Curran1987} As the strain rate is increased, the spall strength of polycrystalline materials can approach that of single crystals due to the fact that the loading pulse length approaches the scale of the distribution of favorable nucleation sites.~\cite{Antoun2003,Wilkerson2016} This behavior is indicative of a transition from energy-limited spall, where there are a large number of failure nucleation sites that allow the sample to fail when it is energetically favorable, to that of flaw-limited spall, where significant elastic energy can be stored within the body before failure sites nucleate.~\cite{Grady1988b} The concept of energy and flaw-limited spall in solids is directly analogous to the concepts of homogeneous and heterogeneous cavitation nucleation in liquids. The comparison of the results from this work and similar studies to those performed at lower strain rates in bullet-piston experiments suggests a transition from energy-limited spall (heterogeneous nucleation) to flaw-limited spall (homogeneous nucleation) as strain rate is increased. It is important to note that this work has focused on the maximum tension supported by the liquid at the onset of cavitation, which represents a small portion of the shock induced cavitation process. A shock wave reflecting from a liquid free surface can produce large cavitation particle clouds and a series of spall layers which break up into a bubbly spray.~\cite{Kolsky1949,Kedrinskii2005} These processes, which occur on much longer timescales than those considered in this work (tens of microseconds to milliseconds), should be expected to be influenced by heterogeneities in the fluid~\cite{Kedrinskii2005}, regardless of the strain rate at the onset of cavitation.

The heterogeneous cavitation theory presented in Section~\ref {heterogeneous} and the discussion above were based on the assumption that the viscosity of the oils during cavitation bubble growth could be adequately represented by the zero-shear-rate viscosity. As was discussed in Section~\ref {materials}, the bubble growth rates required to relieve the tension during plate-impact experiments are likely to induce non-Newtonian behavior in the PDMS fluids, particularly for the intermediate and high-viscosity oils. At shear rates above the critical shear rate ($\dot{\gamma}_\mathrm{c}$), PDMS oils behave as a pseudo-plastic material: their viscosity is shear-rate dependent and they have a resistance to shear deformation (non-zero shear modulus).~\cite{Chhabra2010} While the elasticity of the oils during high strain rate deformation should be expected to increase resistance to bubble growth~\cite{Warnez2015}, the opposite is true of their shear thinning properties, which reduce the apparent viscosity of the fluid. The viscous resistance to bubble growth arises from the extensional flow field as the fluid near the bubble is displaced.~\cite{Pearson1977} In a viscoelastic material, the Newtonian relationship between extensional and shear viscosity ($\eta_\mathrm{E}(\dot{\epsilon})=3\eta(\dot{\gamma})$) is no longer applicable and the ratio between the two values can be much greater than three.~\cite{Jones1987} Although data on the extensional shear rate dependence of PDMS oils is limited, published results of capillary rheometry experiments showed a decay from \SIrange{20}{12}{\pascal\second} as $\dot{\epsilon}$ was increased from \SIrange{e2}{e4}{\per\second} for a 10~St oil, and a decay from \SIrange{125}{25}{\pascal\second} for a 50~St oil over the same range of strain rates.~\cite{Day2008,Williams2010} These results indicate that the high strain rates studied in this work likely affected the viscosity of the high-viscosity silicone oil and reduced the overall difference in the viscosity of the three oils. Nonetheless, it seems reasonable to assume that even at the strain rates experienced in plate-impact experiments (\SI{e4}{\per\second}) there remains a large difference (approximately two orders of magnitude) in the viscosity and resulting resistance to void growth of the three oils considered in this work.

\section{CONCLUSION}
\label{conclusion}
The mechanism by which cavitation nucleates in systems where rupture of the liquid is induced by shock or pressure waves is of significant scientific and engineering interest. The spall strength of three PDMS oils having vastly different viscosities was evaluated over a range of loading conditions using plate-impact experiments. The results showed that the spall strength was unaffected by the variation in the viscosity of the oils or the rate at which tension was applied. These findings were shown to be consistent with a homogeneous cavitation model where the principal mechanism for tension relief is the formation of voids via random molecular fluctuations. The results suggest that cavitation originating from heterogeneities (pre-existing bubbles) does not have a significant effect on the spall strength of the PDMS oils for the conditions probed in this work. Comparison with previously published data obtained using bullet-piston experiments, where the loading rates were two orders of magnitude less than those presented in this work, suggests a transition from homogeneous to heterogeneous cavitation nucleation as the strain rate is increased to that of plate-impact experiments.

\begin{acknowledgments}
The authors would like to thank Jihane Kamil for her help in developing the experiment, Charles Dubois and Asher Bechimol from the Department of Chemical Engineering at École Polytechnique de Montréal for their assistance with the rheometry measurements, Hansen Liu, Zhuo Fan Bao, and Hin Fung Ng for their help in performing experiments, and David Plant and Victoria Suponitsky for their guidance with the experiments. This work was supported by General Fusion and the Natural Sciences and Engineering Research Council of Canada (NSERC) under Collaborative Research and Development Grant 477617-14.
\end{acknowledgments}

\bibliography{bibliography}

\begin{thebibliography}{51}%
\makeatletter
\providecommand \@ifxundefined [1]{%
 \@ifx{#1\undefined}
}%
\providecommand \@ifnum [1]{%
 \ifnum #1\expandafter \@firstoftwo
 \else \expandafter \@secondoftwo
 \fi
}%
\providecommand \@ifx [1]{%
 \ifx #1\expandafter \@firstoftwo
 \else \expandafter \@secondoftwo
 \fi
}%
\providecommand \natexlab [1]{#1}%
\providecommand \enquote  [1]{``#1''}%
\providecommand \bibnamefont  [1]{#1}%
\providecommand \bibfnamefont [1]{#1}%
\providecommand \citenamefont [1]{#1}%
\providecommand \href@noop [0]{\@secondoftwo}%
\providecommand \href [0]{\begingroup \@sanitize@url \@href}%
\providecommand \@href[1]{\@@startlink{#1}\@@href}%
\providecommand \@@href[1]{\endgroup#1\@@endlink}%
\providecommand \@sanitize@url [0]{\catcode `\\12\catcode `\$12\catcode
  `\&12\catcode `\#12\catcode `\^12\catcode `\_12\catcode `\%12\relax}%
\providecommand \@@startlink[1]{}%
\providecommand \@@endlink[0]{}%
\providecommand \url  [0]{\begingroup\@sanitize@url \@url }%
\providecommand \@url [1]{\endgroup\@href {#1}{\urlprefix }}%
\providecommand \urlprefix  [0]{URL }%
\providecommand \Eprint [0]{\href }%
\providecommand \doibase [0]{http://dx.doi.org/}%
\providecommand \selectlanguage [0]{\@gobble}%
\providecommand \bibinfo  [0]{\@secondoftwo}%
\providecommand \bibfield  [0]{\@secondoftwo}%
\providecommand \translation [1]{[#1]}%
\providecommand \BibitemOpen [0]{}%
\providecommand \bibitemStop [0]{}%
\providecommand \bibitemNoStop [0]{.\EOS\space}%
\providecommand \EOS [0]{\spacefactor3000\relax}%
\providecommand \BibitemShut  [1]{\csname bibitem#1\endcsname}%
\let\auto@bib@innerbib\@empty
\bibitem [{\citenamefont {Trevena}(1987)}]{Trevena1987}%
  \BibitemOpen
  \bibfield  {author} {\bibinfo {author} {\bibfnamefont {D.~H.}\ \bibnamefont
  {Trevena}},\ }\href@noop {} {\emph {\bibinfo {title} {Cavitation and
  {T}ension in {L}iquids}}}\ (\bibinfo  {publisher} {Adam Hilger},\ \bibinfo
  {year} {1987})\BibitemShut {NoStop}%
\bibitem [{\citenamefont {Fisher}(1948)}]{Fisher1948}%
  \BibitemOpen
  \bibfield  {author} {\bibinfo {author} {\bibfnamefont {J.~C.}\ \bibnamefont
  {Fisher}},\ }\bibfield  {title} {\enquote {\bibinfo {title} {The fracture of
  liquids},}\ }\href {\doibase 10.1063/1.1698012} {\bibfield  {journal}
  {\bibinfo  {journal} {J. Appl. Phys.}\ }\textbf {\bibinfo {volume} {19}},\
  \bibinfo {pages} {1062--1067} (\bibinfo {year} {1948})}\BibitemShut {NoStop}%
\bibitem [{\citenamefont {Kedrinskii}(2005)}]{Kedrinskii2005}%
  \BibitemOpen
  \bibfield  {author} {\bibinfo {author} {\bibfnamefont {V.~K.}\ \bibnamefont
  {Kedrinskii}},\ }\enquote {\bibinfo {title} {Problems of cavitative
  destruction},}\ in\ \href {\doibase 10.1007/3-540-28563-6_7} {\emph {\bibinfo
  {booktitle} {Hydrodynamics of Explosion: Experiments and Models}}}\ (\bibinfo
   {publisher} {Springer Berlin Heidelberg},\ \bibinfo {year} {2005})\ pp.\
  \bibinfo {pages} {223--296}\BibitemShut {NoStop}%
\bibitem [{\citenamefont {Zeldovich}(1992)}]{Zeldovich1992}%
  \BibitemOpen
  \bibfield  {author} {\bibinfo {author} {\bibfnamefont {Y.~B.}\ \bibnamefont
  {Zeldovich}},\ }\enquote {\bibinfo {title} {On the theory of new phase
  formation: Cavitation},}\ in\ \href@noop {} {\emph {\bibinfo {booktitle}
  {Selected Works of Yakov Borisovich Zeldovich, Volume I: Chemical Physics and
  Hydrodynanics}}},\ \bibinfo {editor} {edited by\ \bibinfo {editor}
  {\bibfnamefont {J.~P.}\ \bibnamefont {Ostriker}}, \bibinfo {editor}
  {\bibfnamefont {G.~I.}\ \bibnamefont {Barenblatt}}, \ and\ \bibinfo {editor}
  {\bibfnamefont {R.~A.}\ \bibnamefont {Sunyaev}}}\ (\bibinfo  {publisher}
  {Princeton University Press},\ \bibinfo {address} {Princeton},\ \bibinfo
  {year} {1992})\ pp.\ \bibinfo {pages} {120--137}\BibitemShut {NoStop}%
\bibitem [{\citenamefont {Carlson}\ and\ \citenamefont
  {Levine}(1975)}]{Carlson1975}%
  \BibitemOpen
  \bibfield  {author} {\bibinfo {author} {\bibfnamefont {G.~A.}\ \bibnamefont
  {Carlson}}\ and\ \bibinfo {author} {\bibfnamefont {H.~S.}\ \bibnamefont
  {Levine}},\ }\bibfield  {title} {\enquote {\bibinfo {title} {Dynamic tensile
  strength of glycerol},}\ }\href {\doibase 10.1063/1.321761} {\bibfield
  {journal} {\bibinfo  {journal} {J. Appl. Phys.}\ }\textbf {\bibinfo {volume}
  {46}},\ \bibinfo {pages} {1594--1601} (\bibinfo {year} {1975})}\BibitemShut
  {NoStop}%
\bibitem [{\citenamefont {Utkin}\ and\ \citenamefont
  {Sosikov}(2009)}]{Utkin2010}%
  \BibitemOpen
  \bibfield  {author} {\bibinfo {author} {\bibfnamefont {A.~V.}\ \bibnamefont
  {Utkin}}\ and\ \bibinfo {author} {\bibfnamefont {V.~A.}\ \bibnamefont
  {Sosikov}},\ }\bibfield  {title} {\enquote {\bibinfo {title} {Tension of
  liquids by shockwaves},}\ }\href {\doibase 10.1063/1.3295202} {\bibfield
  {journal} {\bibinfo  {journal} {AIP Conf. Proc.}\ }\textbf {\bibinfo {volume}
  {1195}},\ \bibinfo {pages} {568--573} (\bibinfo {year} {2009})}\BibitemShut
  {NoStop}%
\bibitem [{\citenamefont {Bull}(1956{\natexlab{a}})}]{Bull1956}%
  \BibitemOpen
  \bibfield  {author} {\bibinfo {author} {\bibfnamefont {T.}~\bibnamefont
  {Bull}},\ }\bibfield  {title} {\enquote {\bibinfo {title} {The tensile
  strengths of viscous liquids under dynamic loading},}\ }\href@noop {}
  {\bibfield  {journal} {\bibinfo  {journal} {Br. J. Appl. Phys.}\ }\textbf
  {\bibinfo {volume} {7}},\ \bibinfo {pages} {416} (\bibinfo {year}
  {1956}{\natexlab{a}})}\BibitemShut {NoStop}%
\bibitem [{\citenamefont {Erlich}, \citenamefont {Wooten},\ and\ \citenamefont
  {Crewdson}(1971)}]{Erlich1971}%
  \BibitemOpen
  \bibfield  {author} {\bibinfo {author} {\bibfnamefont {D.~C.}\ \bibnamefont
  {Erlich}}, \bibinfo {author} {\bibfnamefont {D.~C.}\ \bibnamefont {Wooten}},
  \ and\ \bibinfo {author} {\bibfnamefont {R.~C.}\ \bibnamefont {Crewdson}},\
  }\bibfield  {title} {\enquote {\bibinfo {title} {Dynamic tensile failure of
  glycerol},}\ }\href {\doibase 10.1063/1.1659970} {\bibfield  {journal}
  {\bibinfo  {journal} {J. Appl. Phys.}\ }\textbf {\bibinfo {volume} {42}},\
  \bibinfo {pages} {5495--5502} (\bibinfo {year} {1971})}\BibitemShut {NoStop}%
\bibitem [{\citenamefont {Kolsky}\ \emph {et~al.}(1949)\citenamefont {Kolsky},
  \citenamefont {Lewis}, \citenamefont {Sampson}, \citenamefont {Shearman},
  \citenamefont {Snow},\ and\ \citenamefont {Taylor}}]{Kolsky1949}%
  \BibitemOpen
  \bibfield  {author} {\bibinfo {author} {\bibfnamefont {H.}~\bibnamefont
  {Kolsky}}, \bibinfo {author} {\bibfnamefont {J.~P.}\ \bibnamefont {Lewis}},
  \bibinfo {author} {\bibfnamefont {M.~T.}\ \bibnamefont {Sampson}}, \bibinfo
  {author} {\bibfnamefont {A.~C.}\ \bibnamefont {Shearman}}, \bibinfo {author}
  {\bibfnamefont {C.~I.}\ \bibnamefont {Snow}}, \ and\ \bibinfo {author}
  {\bibfnamefont {G.~I.}\ \bibnamefont {Taylor}},\ }\bibfield  {title}
  {\enquote {\bibinfo {title} {Splashes from underwater explosions},}\ }\href
  {\doibase 10.1098/rspa.1949.0034} {\bibfield  {journal} {\bibinfo  {journal}
  {Proc. Royal Soc. Lond. A}\ }\textbf {\bibinfo {volume} {196}},\ \bibinfo
  {pages} {379--402} (\bibinfo {year} {1949})}\BibitemShut {NoStop}%
\bibitem [{\citenamefont {Cole}(1948)}]{Cole1948}%
  \BibitemOpen
  \bibfield  {author} {\bibinfo {author} {\bibfnamefont {R.}~\bibnamefont
  {Cole}},\ }\href@noop {} {\emph {\bibinfo {title} {{Underwater
  Explosions}}}}\ (\bibinfo  {publisher} {Princeton University Press},\
  \bibinfo {address} {Princeton},\ \bibinfo {year} {1948})\BibitemShut
  {NoStop}%
\bibitem [{\citenamefont {Zhu}\ \emph {et~al.}(2002)\citenamefont {Zhu},
  \citenamefont {Cocks}, \citenamefont {Preminger},\ and\ \citenamefont
  {Zhong}}]{Zhu2002}%
  \BibitemOpen
  \bibfield  {author} {\bibinfo {author} {\bibfnamefont {S.}~\bibnamefont
  {Zhu}}, \bibinfo {author} {\bibfnamefont {F.~H.}\ \bibnamefont {Cocks}},
  \bibinfo {author} {\bibfnamefont {G.~M.}\ \bibnamefont {Preminger}}, \ and\
  \bibinfo {author} {\bibfnamefont {P.}~\bibnamefont {Zhong}},\ }\bibfield
  {title} {\enquote {\bibinfo {title} {The role of stress waves and cavitation
  in stone comminution in shock wave lithotripsy},}\ }\href {\doibase
  https://doi.org/10.1016/S0301-5629(02)00506-9} {\bibfield  {journal}
  {\bibinfo  {journal} {Ultrasound Med. Biol.}\ }\textbf {\bibinfo {volume}
  {28}},\ \bibinfo {pages} {661 -- 671} (\bibinfo {year} {2002})}\BibitemShut
  {NoStop}%
\bibitem [{\citenamefont {Milne}\ \emph {et~al.}(2017)\citenamefont {Milne},
  \citenamefont {Longbottom}, \citenamefont {Frost}, \citenamefont {Loiseau},
  \citenamefont {Goroshin},\ and\ \citenamefont {Petel}}]{Milne2017}%
  \BibitemOpen
  \bibfield  {author} {\bibinfo {author} {\bibfnamefont {A.}~\bibnamefont
  {Milne}}, \bibinfo {author} {\bibfnamefont {A.}~\bibnamefont {Longbottom}},
  \bibinfo {author} {\bibfnamefont {D.~L.}\ \bibnamefont {Frost}}, \bibinfo
  {author} {\bibfnamefont {J.}~\bibnamefont {Loiseau}}, \bibinfo {author}
  {\bibfnamefont {S.}~\bibnamefont {Goroshin}}, \ and\ \bibinfo {author}
  {\bibfnamefont {O.}~\bibnamefont {Petel}},\ }\bibfield  {title} {\enquote
  {\bibinfo {title} {Explosive fragmentation of liquids in spherical
  geometry},}\ }\href {\doibase 10.1007/s00193-016-0671-y} {\bibfield
  {journal} {\bibinfo  {journal} {Shock Waves}\ }\textbf {\bibinfo {volume}
  {27}},\ \bibinfo {pages} {383--393} (\bibinfo {year} {2017})}\BibitemShut
  {NoStop}%
\bibitem [{\citenamefont {Laberge}(2008)}]{Laberge2008}%
  \BibitemOpen
  \bibfield  {author} {\bibinfo {author} {\bibfnamefont {M.}~\bibnamefont
  {Laberge}},\ }\bibfield  {title} {\enquote {\bibinfo {title} {An acoustically
  driven magnetized target fusion reactor},}\ }\href {\doibase
  10.1007/s10894-007-9091-4} {\bibfield  {journal} {\bibinfo  {journal} {J.
  Fusion Energy}\ }\textbf {\bibinfo {volume} {27}},\ \bibinfo {pages} {65--68}
  (\bibinfo {year} {2008})}\BibitemShut {NoStop}%
\bibitem [{\citenamefont {Laberge}(2009)}]{Laberge2009}%
  \BibitemOpen
  \bibfield  {author} {\bibinfo {author} {\bibfnamefont {M.}~\bibnamefont
  {Laberge}},\ }\bibfield  {title} {\enquote {\bibinfo {title} {Experimental
  results for an acoustic driver for {M}{T}{F}},}\ }\href {\doibase
  10.1007/s10894-008-9181-y} {\bibfield  {journal} {\bibinfo  {journal} {J.
  Fusion Energy}\ }\textbf {\bibinfo {volume} {28}},\ \bibinfo {pages}
  {179--182} (\bibinfo {year} {2009})}\BibitemShut {NoStop}%
\bibitem [{\citenamefont {Suponitsky}, \citenamefont {Froese},\ and\
  \citenamefont {Barsky}(2014)}]{Suponitsky2014}%
  \BibitemOpen
  \bibfield  {author} {\bibinfo {author} {\bibfnamefont {V.}~\bibnamefont
  {Suponitsky}}, \bibinfo {author} {\bibfnamefont {A.}~\bibnamefont {Froese}},
  \ and\ \bibinfo {author} {\bibfnamefont {S.}~\bibnamefont {Barsky}},\
  }\bibfield  {title} {\enquote {\bibinfo {title} {{R}ichtmyer--{M}eshkov
  instability of a liquid–gas interface driven by a cylindrical imploding
  pressure wave},}\ }\href {\doibase
  https://doi.org/10.1016/j.compfluid.2013.10.031} {\bibfield  {journal}
  {\bibinfo  {journal} {Comput. Fluids}\ }\textbf {\bibinfo {volume} {89}},\
  \bibinfo {pages} {1 -- 19} (\bibinfo {year} {2014})}\BibitemShut {NoStop}%
\bibitem [{\citenamefont {Suponitsky}\ \emph {et~al.}(2017)\citenamefont
  {Suponitsky}, \citenamefont {Plant}, \citenamefont {Avital},\ and\
  \citenamefont {Munjiza}}]{Suponitsky2017}%
  \BibitemOpen
  \bibfield  {author} {\bibinfo {author} {\bibfnamefont {V.}~\bibnamefont
  {Suponitsky}}, \bibinfo {author} {\bibfnamefont {D.}~\bibnamefont {Plant}},
  \bibinfo {author} {\bibfnamefont {E.~J.}\ \bibnamefont {Avital}}, \ and\
  \bibinfo {author} {\bibfnamefont {A.}~\bibnamefont {Munjiza}},\ }\bibfield
  {title} {\enquote {\bibinfo {title} {Pressure wave in liquid generated by
  pneumatic pistons and its interaction with a free surface},}\ }\href
  {\doibase 10.1142/S1758825117500375} {\bibfield  {journal} {\bibinfo
  {journal} {Int. J. Appl. Mech.}\ }\textbf {\bibinfo {volume} {09}},\ \bibinfo
  {pages} {1750037} (\bibinfo {year} {2017})}\BibitemShut {NoStop}%
\bibitem [{\citenamefont {Antoun}\ \emph {et~al.}(2003)\citenamefont {Antoun},
  \citenamefont {Seaman}, \citenamefont {Curran}, \citenamefont {Kanel},
  \citenamefont {Razorenov},\ and\ \citenamefont {Utkin}}]{Antoun2003}%
  \BibitemOpen
  \bibfield  {author} {\bibinfo {author} {\bibfnamefont {T.}~\bibnamefont
  {Antoun}}, \bibinfo {author} {\bibfnamefont {L.}~\bibnamefont {Seaman}},
  \bibinfo {author} {\bibfnamefont {D.~R.}\ \bibnamefont {Curran}}, \bibinfo
  {author} {\bibfnamefont {G.~I.}\ \bibnamefont {Kanel}}, \bibinfo {author}
  {\bibfnamefont {S.~V.}\ \bibnamefont {Razorenov}}, \ and\ \bibinfo {author}
  {\bibfnamefont {A.~V.}\ \bibnamefont {Utkin}},\ }\href@noop {} {\emph
  {\bibinfo {title} {{Spall Fracture}}}},\ \bibinfo {edition} {1st}\ ed.,\
  edited by\ \bibinfo {editor} {\bibfnamefont {L.}~\bibnamefont {Davidson}}\
  and\ \bibinfo {editor} {\bibfnamefont {Y.}~\bibnamefont {Horie}}\ (\bibinfo
  {publisher} {Springer-Verlag},\ \bibinfo {address} {New York},\ \bibinfo
  {year} {2003})\BibitemShut {NoStop}%
\bibitem [{\citenamefont {Couzens}\ and\ \citenamefont
  {Trevena}(1974)}]{Couzens1974}%
  \BibitemOpen
  \bibfield  {author} {\bibinfo {author} {\bibfnamefont {D.~C.~F.}\
  \bibnamefont {Couzens}}\ and\ \bibinfo {author} {\bibfnamefont {D.~H.}\
  \bibnamefont {Trevena}},\ }\bibfield  {title} {\enquote {\bibinfo {title}
  {Tensile failure of liquids under dynamic stressing},}\ }\href
  {http://stacks.iop.org/0022-3727/7/i=16/a=315} {\bibfield  {journal}
  {\bibinfo  {journal} {J. Phys. D}\ }\textbf {\bibinfo {volume} {7}},\
  \bibinfo {pages} {2277} (\bibinfo {year} {1974})}\BibitemShut {NoStop}%
\bibitem [{\citenamefont {Bull}(1956{\natexlab{b}})}]{Bull1956b}%
  \BibitemOpen
  \bibfield  {author} {\bibinfo {author} {\bibfnamefont {T.~H.}\ \bibnamefont
  {Bull}},\ }\bibfield  {title} {\enquote {\bibinfo {title} {Xiii. the tensile
  strengths of liquids under dynamic loading},}\ }\href {\doibase
  10.1080/14786435608238088} {\bibfield  {journal} {\bibinfo  {journal}
  {Philos. Mag.}\ }\textbf {\bibinfo {volume} {1}},\ \bibinfo {pages}
  {153--165} (\bibinfo {year} {1956}{\natexlab{b}})}\BibitemShut {NoStop}%
\bibitem [{\citenamefont {Grady}(1988{\natexlab{a}})}]{Grady1988}%
  \BibitemOpen
  \bibfield  {author} {\bibinfo {author} {\bibfnamefont {D.~E.}\ \bibnamefont
  {Grady}},\ }\bibfield  {title} {\enquote {\bibinfo {title} {{The spall
  strength of condensed matter}},}\ }\href@noop {} {\bibfield  {journal}
  {\bibinfo  {journal} {J. Mech. Phys. Solids}\ }\textbf {\bibinfo {volume}
  {36}},\ \bibinfo {pages} {353--384} (\bibinfo {year}
  {1988}{\natexlab{a}})}\BibitemShut {NoStop}%
\bibitem [{\citenamefont {Huneault}\ \emph {et~al.}(2018)\citenamefont
  {Huneault}, \citenamefont {Kamil}, \citenamefont {Higgins},\ and\
  \citenamefont {Plant}}]{Huneault2018}%
  \BibitemOpen
  \bibfield  {author} {\bibinfo {author} {\bibfnamefont {J.}~\bibnamefont
  {Huneault}}, \bibinfo {author} {\bibfnamefont {J.}~\bibnamefont {Kamil}},
  \bibinfo {author} {\bibfnamefont {A.}~\bibnamefont {Higgins}}, \ and\
  \bibinfo {author} {\bibfnamefont {D.}~\bibnamefont {Plant}},\ }\bibfield
  {title} {\enquote {\bibinfo {title} {Dynamic tensile strength of silicone
  oils},}\ }\href {\doibase 10.1063/1.5044825} {\bibfield  {journal} {\bibinfo
  {journal} {AIP Conf. Proc.}\ }\textbf {\bibinfo {volume} {1979}},\ \bibinfo
  {pages} {070016} (\bibinfo {year} {2018})}\BibitemShut {NoStop}%
\bibitem [{\citenamefont {Patterson}(1998)}]{Patterson1998}%
  \BibitemOpen
  \bibfield  {author} {\bibinfo {author} {\bibfnamefont {R.~F.}\ \bibnamefont
  {Patterson}},\ }\enquote {\bibinfo {title} {9 - silicones},}\ in\ \href
  {\doibase https://doi.org/10.1016/B978-081551421-3.50012-6} {\emph {\bibinfo
  {booktitle} {Handbook of Thermoset Plastics}}},\ \bibinfo {editor} {edited
  by\ \bibinfo {editor} {\bibfnamefont {S.~H.}\ \bibnamefont {Goodman}}}\
  (\bibinfo  {publisher} {William Andrew Publishing},\ \bibinfo {address}
  {Westwood, NJ},\ \bibinfo {year} {1998})\ pp.\ \bibinfo {pages} {468 --
  497},\ \bibinfo {edition} {2nd}\ ed.\BibitemShut {Stop}%
\bibitem [{\citenamefont {Leibacher}, \citenamefont {Reichert},\ and\
  \citenamefont {Dual}(2015)}]{Leibacher2015}%
  \BibitemOpen
  \bibfield  {author} {\bibinfo {author} {\bibfnamefont {I.}~\bibnamefont
  {Leibacher}}, \bibinfo {author} {\bibfnamefont {P.}~\bibnamefont {Reichert}},
  \ and\ \bibinfo {author} {\bibfnamefont {J.}~\bibnamefont {Dual}},\
  }\bibfield  {title} {\enquote {\bibinfo {title} {Microfluidic droplet
  handling by bulk acoustic wave ({BAW}) acoustophoresis},}\ }\href {\doibase
  10.1039/C5LC00083A} {\bibfield  {journal} {\bibinfo  {journal} {Lab Chip}\
  }\textbf {\bibinfo {volume} {15}},\ \bibinfo {pages} {2896--2905} (\bibinfo
  {year} {2015})}\BibitemShut {NoStop}%
\bibitem [{\citenamefont {Barlow}\ \emph {et~al.}(1964)\citenamefont {Barlow},
  \citenamefont {Harrison}, \citenamefont {Lamb},\ and\ \citenamefont
  {Robertson}}]{Barlow1964}%
  \BibitemOpen
  \bibfield  {author} {\bibinfo {author} {\bibfnamefont {A.~J.}\ \bibnamefont
  {Barlow}}, \bibinfo {author} {\bibfnamefont {G.}~\bibnamefont {Harrison}},
  \bibinfo {author} {\bibfnamefont {J.}~\bibnamefont {Lamb}}, \ and\ \bibinfo
  {author} {\bibfnamefont {J.~M.}\ \bibnamefont {Robertson}},\ }\bibfield
  {title} {\enquote {\bibinfo {title} {Viscoelastic relaxation of
  polydimethylsiloxane liquids},}\ }\href {\doibase 10.1098/rspa.1964.0229}
  {\bibfield  {journal} {\bibinfo  {journal} {Proc. Royal Soc. Lond. A}\
  }\textbf {\bibinfo {volume} {282}},\ \bibinfo {pages} {228--251} (\bibinfo
  {year} {1964})}\BibitemShut {NoStop}%
\bibitem [{\citenamefont {Ghannam}\ and\ \citenamefont
  {Esmail}(1998)}]{Ghannam1998}%
  \BibitemOpen
  \bibfield  {author} {\bibinfo {author} {\bibfnamefont {M.~T.}\ \bibnamefont
  {Ghannam}}\ and\ \bibinfo {author} {\bibfnamefont {M.~N.}\ \bibnamefont
  {Esmail}},\ }\bibfield  {title} {\enquote {\bibinfo {title} {Rheological
  properties of poly(dimethylsiloxane)},}\ }\href {\doibase 10.1021/ie9703346}
  {\bibfield  {journal} {\bibinfo  {journal} {Ind. Eng. Chem. Res.}\ }\textbf
  {\bibinfo {volume} {37}},\ \bibinfo {pages} {1335--1340} (\bibinfo {year}
  {1998})}\BibitemShut {NoStop}%
\bibitem [{\citenamefont {Carré}\ and\ \citenamefont
  {Woehl}(2006)}]{Carre2006}%
  \BibitemOpen
  \bibfield  {author} {\bibinfo {author} {\bibfnamefont {A.}~\bibnamefont
  {Carré}}\ and\ \bibinfo {author} {\bibfnamefont {P.}~\bibnamefont {Woehl}},\
  }\bibfield  {title} {\enquote {\bibinfo {title} {Spreading of silicone oils
  on glass in two geometries},}\ }\href {\doibase 10.1021/la0518997} {\bibfield
   {journal} {\bibinfo  {journal} {Langmuir}\ }\textbf {\bibinfo {volume}
  {22}},\ \bibinfo {pages} {134--139} (\bibinfo {year} {2006})},\ \bibinfo
  {note} {pMID: 16378411}\BibitemShut {NoStop}%
\bibitem [{\citenamefont {Vázquez-Quesada}\ \emph {et~al.}(2017)\citenamefont
  {Vázquez-Quesada}, \citenamefont {Mahmud}, \citenamefont {Dai},
  \citenamefont {Ellero},\ and\ \citenamefont {Tanner}}]{Vazquez2017}%
  \BibitemOpen
  \bibfield  {author} {\bibinfo {author} {\bibfnamefont {A.}~\bibnamefont
  {Vázquez-Quesada}}, \bibinfo {author} {\bibfnamefont {A.}~\bibnamefont
  {Mahmud}}, \bibinfo {author} {\bibfnamefont {S.}~\bibnamefont {Dai}},
  \bibinfo {author} {\bibfnamefont {M.}~\bibnamefont {Ellero}}, \ and\ \bibinfo
  {author} {\bibfnamefont {R.~I.}\ \bibnamefont {Tanner}},\ }\bibfield  {title}
  {\enquote {\bibinfo {title} {Investigating the causes of shear-thinning in
  non-colloidal suspensions: Experiments and simulations},}\ }\href {\doibase
  https://doi.org/10.1016/j.jnnfm.2017.08.005} {\bibfield  {journal} {\bibinfo
  {journal} {J. Non-Newt. Fluid Mech.}\ }\textbf {\bibinfo {volume} {248}},\
  \bibinfo {pages} {1 -- 7} (\bibinfo {year} {2017})}\BibitemShut {NoStop}%
\bibitem [{\citenamefont {Strand}\ \emph {et~al.}(2006)\citenamefont {Strand},
  \citenamefont {Goosman}, \citenamefont {Martinez}, \citenamefont
  {Whitworth},\ and\ \citenamefont {Kuhlow}}]{Strand2006}%
  \BibitemOpen
  \bibfield  {author} {\bibinfo {author} {\bibfnamefont {O.~T.}\ \bibnamefont
  {Strand}}, \bibinfo {author} {\bibfnamefont {D.~R.}\ \bibnamefont {Goosman}},
  \bibinfo {author} {\bibfnamefont {C.}~\bibnamefont {Martinez}}, \bibinfo
  {author} {\bibfnamefont {T.~L.}\ \bibnamefont {Whitworth}}, \ and\ \bibinfo
  {author} {\bibfnamefont {W.~W.}\ \bibnamefont {Kuhlow}},\ }\bibfield  {title}
  {\enquote {\bibinfo {title} {Compact system for high-speed velocimetry using
  heterodyne techniques},}\ }\href {\doibase 10.1063/1.2336749} {\bibfield
  {journal} {\bibinfo  {journal} {Rev. Sci. Instrum.}\ }\textbf {\bibinfo
  {volume} {77}},\ \bibinfo {pages} {083108} (\bibinfo {year}
  {2006})}\BibitemShut {NoStop}%
\bibitem [{\citenamefont {Dolan}(2010)}]{Dolan2010}%
  \BibitemOpen
  \bibfield  {author} {\bibinfo {author} {\bibfnamefont {D.~H.}\ \bibnamefont
  {Dolan}},\ }\bibfield  {title} {\enquote {\bibinfo {title} {Accuracy and
  precision in photonic doppler velocimetry},}\ }\href {\doibase
  10.1063/1.3429257} {\bibfield  {journal} {\bibinfo  {journal} {Rev. Sci.
  Instrum.}\ }\textbf {\bibinfo {volume} {81}},\ \bibinfo {pages} {053905}
  (\bibinfo {year} {2010})}\BibitemShut {NoStop}%
\bibitem [{\citenamefont {Luo}\ \emph {et~al.}(2009)\citenamefont {Luo},
  \citenamefont {An}, \citenamefont {Germann},\ and\ \citenamefont
  {Han}}]{Sheng2009}%
  \BibitemOpen
  \bibfield  {author} {\bibinfo {author} {\bibfnamefont {S.-N.}\ \bibnamefont
  {Luo}}, \bibinfo {author} {\bibfnamefont {Q.}~\bibnamefont {An}}, \bibinfo
  {author} {\bibfnamefont {T.~C.}\ \bibnamefont {Germann}}, \ and\ \bibinfo
  {author} {\bibfnamefont {L.-B.}\ \bibnamefont {Han}},\ }\bibfield  {title}
  {\enquote {\bibinfo {title} {Shock-induced spall in solid and liquid cu at
  extreme strain rates},}\ }\href {\doibase 10.1063/1.3158062} {\bibfield
  {journal} {\bibinfo  {journal} {J. Appl. Phys.}\ }\textbf {\bibinfo {volume}
  {106}},\ \bibinfo {pages} {013502} (\bibinfo {year} {2009})}\BibitemShut
  {NoStop}%
\bibitem [{\citenamefont {Walsh}\ and\ \citenamefont
  {Christian}(1955)}]{Walsh1955}%
  \BibitemOpen
  \bibfield  {author} {\bibinfo {author} {\bibfnamefont {J.~M.}\ \bibnamefont
  {Walsh}}\ and\ \bibinfo {author} {\bibfnamefont {R.~H.}\ \bibnamefont
  {Christian}},\ }\bibfield  {title} {\enquote {\bibinfo {title} {Equation of
  state of metals from shock wave measurements},}\ }\href {\doibase
  10.1103/PhysRev.97.1544} {\bibfield  {journal} {\bibinfo  {journal} {Phys.
  Rev.}\ }\textbf {\bibinfo {volume} {97}},\ \bibinfo {pages} {1544--1556}
  (\bibinfo {year} {1955})}\BibitemShut {NoStop}%
\bibitem [{\citenamefont {Cooper}(2002)}]{Cooper2002}%
  \BibitemOpen
  \bibfield  {author} {\bibinfo {author} {\bibfnamefont {P.~W.}\ \bibnamefont
  {Cooper}},\ }\href@noop {} {\emph {\bibinfo {title} {Explosives
  {E}ngineering}}},\ \bibinfo {edition} {4th}\ ed.\ (\bibinfo  {publisher}
  {Wiley-VCH},\ \bibinfo {address} {New York},\ \bibinfo {year}
  {2002})\BibitemShut {NoStop}%
\bibitem [{\citenamefont {Woolfolk}, \citenamefont {Cowperthwaite},\ and\
  \citenamefont {Shaw}(1973)}]{Woolfolk1973}%
  \BibitemOpen
  \bibfield  {author} {\bibinfo {author} {\bibfnamefont {R.}~\bibnamefont
  {Woolfolk}}, \bibinfo {author} {\bibfnamefont {M.}~\bibnamefont
  {Cowperthwaite}}, \ and\ \bibinfo {author} {\bibfnamefont {R.}~\bibnamefont
  {Shaw}},\ }\bibfield  {title} {\enquote {\bibinfo {title} {A “universal”
  {H}ugoniot for liquids},}\ }\href {\doibase
  https://doi.org/10.1016/0040-6031(73)80019-X} {\bibfield  {journal} {\bibinfo
   {journal} {Thermochim Acta}\ }\textbf {\bibinfo {volume} {5}},\ \bibinfo
  {pages} {409 -- 414} (\bibinfo {year} {1973})}\BibitemShut {NoStop}%
\bibitem [{\citenamefont {Garrett}, \citenamefont {Chhabildas},\ and\
  \citenamefont {Reinhart}(2006)}]{Garrett2006}%
  \BibitemOpen
  \bibfield  {author} {\bibinfo {author} {\bibfnamefont {G.~R.}\ \bibnamefont
  {Garrett}}, \bibinfo {author} {\bibfnamefont {L.~C.}\ \bibnamefont
  {Chhabildas}}, \ and\ \bibinfo {author} {\bibfnamefont {W.~D.}\ \bibnamefont
  {Reinhart}},\ }\bibfield  {title} {\enquote {\bibinfo {title} {Shock
  compression of liquids},}\ }\href {\doibase 10.1063/1.2263270} {\bibfield
  {journal} {\bibinfo  {journal} {AIP Conf. Proc.}\ }\textbf {\bibinfo {volume}
  {845}},\ \bibinfo {pages} {81--84} (\bibinfo {year} {2006})}\BibitemShut
  {NoStop}%
\bibitem [{\citenamefont {Utkin}(1997)}]{Utkin1997}%
  \BibitemOpen
  \bibfield  {author} {\bibinfo {author} {\bibfnamefont {A.~V.}\ \bibnamefont
  {Utkin}},\ }\bibfield  {title} {\enquote {\bibinfo {title} {Determination of
  the constants of spall-fracture kinetics of materials using experimental
  data},}\ }\href {\doibase 10.1007/BF02467962} {\bibfield  {journal} {\bibinfo
   {journal} {J. Appl. Mech. Tech. Phys.}\ }\textbf {\bibinfo {volume} {38}},\
  \bibinfo {pages} {952--960} (\bibinfo {year} {1997})}\BibitemShut {NoStop}%
\bibitem [{\citenamefont {Utkin}, \citenamefont {Sosikov},\ and\ \citenamefont
  {Fortov}(2006)}]{Utkin2006}%
  \BibitemOpen
  \bibfield  {author} {\bibinfo {author} {\bibfnamefont {A.~V.}\ \bibnamefont
  {Utkin}}, \bibinfo {author} {\bibfnamefont {V.~A.}\ \bibnamefont {Sosikov}},
  \ and\ \bibinfo {author} {\bibfnamefont {V.~E.}\ \bibnamefont {Fortov}},\
  }\bibfield  {title} {\enquote {\bibinfo {title} {Tension of ethyl alcohol and
  hexadecane by shock waves},}\ }\href {\doibase 10.1063/1.2263466} {\bibfield
  {journal} {\bibinfo  {journal} {AIP Conf. Proc.}\ }\textbf {\bibinfo {volume}
  {845}},\ \bibinfo {pages} {896--899} (\bibinfo {year} {2006})}\BibitemShut
  {NoStop}%
\bibitem [{\citenamefont {Bogach}\ and\ \citenamefont
  {Utkin}(2000)}]{Bogach2000}%
  \BibitemOpen
  \bibfield  {author} {\bibinfo {author} {\bibfnamefont {A.~A.}\ \bibnamefont
  {Bogach}}\ and\ \bibinfo {author} {\bibfnamefont {A.~V.}\ \bibnamefont
  {Utkin}},\ }\bibfield  {title} {\enquote {\bibinfo {title} {Strength of water
  under pulsed loading},}\ }\href {\doibase 10.1007/BF02466877} {\bibfield
  {journal} {\bibinfo  {journal} {J. Appl. Mech. Tech. Phys.}\ }\textbf
  {\bibinfo {volume} {41}},\ \bibinfo {pages} {752--758} (\bibinfo {year}
  {2000})}\BibitemShut {NoStop}%
\bibitem [{\citenamefont {Plesset}\ and\ \citenamefont
  {Prosperetti}(1977)}]{Plesset1977}%
  \BibitemOpen
  \bibfield  {author} {\bibinfo {author} {\bibfnamefont {M.~S.}\ \bibnamefont
  {Plesset}}\ and\ \bibinfo {author} {\bibfnamefont {A.}~\bibnamefont
  {Prosperetti}},\ }\bibfield  {title} {\enquote {\bibinfo {title} {Bubble
  dynamics and cavitation},}\ }\href {\doibase
  10.1146/annurev.fl.09.010177.001045} {\bibfield  {journal} {\bibinfo
  {journal} {Annu. Rev. Fluid Mech.}\ }\textbf {\bibinfo {volume} {9}},\
  \bibinfo {pages} {145--185} (\bibinfo {year} {1977})}\BibitemShut {NoStop}%
\bibitem [{\citenamefont {Couzens}\ and\ \citenamefont
  {Trevena}(1969)}]{Couzens1969}%
  \BibitemOpen
  \bibfield  {author} {\bibinfo {author} {\bibfnamefont {D.}~\bibnamefont
  {Couzens}}\ and\ \bibinfo {author} {\bibfnamefont {D.}~\bibnamefont
  {Trevena}},\ }\bibfield  {title} {\enquote {\bibinfo {title} {Critical
  tension in a liquid under dynamic conditions of stressing},}\ }\href
  {\doibase 10.1038/222473a0} {\bibfield  {journal} {\bibinfo  {journal}
  {Nature}\ }\textbf {\bibinfo {volume} {222}},\ \bibinfo {pages} {473}
  (\bibinfo {year} {1969})}\BibitemShut {NoStop}%
\bibitem [{\citenamefont {Utkin}, \citenamefont {Sosikov},\ and\ \citenamefont
  {Bogach}(2003)}]{Utkin2003}%
  \BibitemOpen
  \bibfield  {author} {\bibinfo {author} {\bibfnamefont {A.~V.}\ \bibnamefont
  {Utkin}}, \bibinfo {author} {\bibfnamefont {V.~A.}\ \bibnamefont {Sosikov}},
  \ and\ \bibinfo {author} {\bibfnamefont {A.~A.}\ \bibnamefont {Bogach}},\
  }\bibfield  {title} {\enquote {\bibinfo {title} {Impulsive tension of hexane
  and glycerol under shock‐wave loading},}\ }\href {\doibase
  10.1023/A:1022532107673} {\bibfield  {journal} {\bibinfo  {journal} {J. Appl.
  Mech. Tech. Phys.}\ }\textbf {\bibinfo {volume} {44}},\ \bibinfo {pages}
  {174--179} (\bibinfo {year} {2003})}\BibitemShut {NoStop}%
\bibitem [{\citenamefont {Curran}, \citenamefont {Seaman},\ and\ \citenamefont
  {Shockey}(1987)}]{Curran1987}%
  \BibitemOpen
  \bibfield  {author} {\bibinfo {author} {\bibfnamefont {D.}~\bibnamefont
  {Curran}}, \bibinfo {author} {\bibfnamefont {L.}~\bibnamefont {Seaman}}, \
  and\ \bibinfo {author} {\bibfnamefont {D.}~\bibnamefont {Shockey}},\
  }\bibfield  {title} {\enquote {\bibinfo {title} {Dynamic failure of
  solids},}\ }\href {\doibase https://doi.org/10.1016/0370-1573(87)90049-4}
  {\bibfield  {journal} {\bibinfo  {journal} {Phys. Rep.}\ }\textbf {\bibinfo
  {volume} {147}},\ \bibinfo {pages} {253 -- 388} (\bibinfo {year}
  {1987})}\BibitemShut {NoStop}%
\bibitem [{\citenamefont {Razorenov}\ \emph {et~al.}(2006)\citenamefont
  {Razorenov}, \citenamefont {Kanel}, \citenamefont {Savinykh},\ and\
  \citenamefont {Fortov}}]{Razorenov2006}%
  \BibitemOpen
  \bibfield  {author} {\bibinfo {author} {\bibfnamefont {S.~V.}\ \bibnamefont
  {Razorenov}}, \bibinfo {author} {\bibfnamefont {G.~I.}\ \bibnamefont
  {Kanel}}, \bibinfo {author} {\bibfnamefont {A.~S.}\ \bibnamefont {Savinykh}},
  \ and\ \bibinfo {author} {\bibfnamefont {V.~E.}\ \bibnamefont {Fortov}},\
  }\bibfield  {title} {\enquote {\bibinfo {title} {Large tensions and strength
  of iron in different structure states},}\ }\href {\doibase 10.1063/1.2263406}
  {\bibfield  {journal} {\bibinfo  {journal} {AIP Conf. Proc.}\ }\textbf
  {\bibinfo {volume} {845}},\ \bibinfo {pages} {650--653} (\bibinfo {year}
  {2006})}\BibitemShut {NoStop}%
\bibitem [{\citenamefont {Razorenov}\ \emph {et~al.}(2007)\citenamefont
  {Razorenov}, \citenamefont {Kanel}, \citenamefont {Herrmann}, \citenamefont
  {Zaretsky},\ and\ \citenamefont {Ivanchihina}}]{Razorenov2007}%
  \BibitemOpen
  \bibfield  {author} {\bibinfo {author} {\bibfnamefont {S.~V.}\ \bibnamefont
  {Razorenov}}, \bibinfo {author} {\bibfnamefont {G.~I.}\ \bibnamefont
  {Kanel}}, \bibinfo {author} {\bibfnamefont {B.}~\bibnamefont {Herrmann}},
  \bibinfo {author} {\bibfnamefont {E.~B.}\ \bibnamefont {Zaretsky}}, \ and\
  \bibinfo {author} {\bibfnamefont {G.~E.}\ \bibnamefont {Ivanchihina}},\
  }\bibfield  {title} {\enquote {\bibinfo {title} {Influence of nano‐size
  inclusions on spall fracture of copper single crystals},}\ }\href {\doibase
  10.1063/1.2833154} {\bibfield  {journal} {\bibinfo  {journal} {AIP Conf.
  Proc.}\ }\textbf {\bibinfo {volume} {955}},\ \bibinfo {pages} {581--584}
  (\bibinfo {year} {2007})}\BibitemShut {NoStop}%
\bibitem [{\citenamefont {Wilkerson}\ and\ \citenamefont
  {Ramesh}(2016)}]{Wilkerson2016}%
  \BibitemOpen
  \bibfield  {author} {\bibinfo {author} {\bibfnamefont {J.~W.}\ \bibnamefont
  {Wilkerson}}\ and\ \bibinfo {author} {\bibfnamefont {K.~T.}\ \bibnamefont
  {Ramesh}},\ }\bibfield  {title} {\enquote {\bibinfo {title} {Unraveling the
  anomalous grain size dependence of cavitation},}\ }\href {\doibase
  10.1103/PhysRevLett.117.215503} {\bibfield  {journal} {\bibinfo  {journal}
  {Phys. Rev. Lett.}\ }\textbf {\bibinfo {volume} {117}},\ \bibinfo {pages}
  {215503} (\bibinfo {year} {2016})}\BibitemShut {NoStop}%
\bibitem [{\citenamefont {Grady}(1988{\natexlab{b}})}]{Grady1988b}%
  \BibitemOpen
  \bibfield  {author} {\bibinfo {author} {\bibfnamefont {D.}~\bibnamefont
  {Grady}},\ }\bibfield  {title} {\enquote {\bibinfo {title} {Incipient spall,
  crack branching, and fragmentation statistics in the spall process},}\ }\href
  {\doibase 10.1051/jphyscol:1988326} {\bibfield  {journal} {\bibinfo
  {journal} {J. Phys. Colloq.}\ }\textbf {\bibinfo {volume} {49}},\ \bibinfo
  {pages} {C3--175} (\bibinfo {year} {1988}{\natexlab{b}})}\BibitemShut
  {NoStop}%
\bibitem [{\citenamefont {Chhabra}(2010)}]{Chhabra2010}%
  \BibitemOpen
  \bibfield  {author} {\bibinfo {author} {\bibfnamefont {R.~P.}\ \bibnamefont
  {Chhabra}},\ }\enquote {\bibinfo {title} {Non-newtonian fluids: An
  introduction},}\ in\ \href {\doibase 10.1007/978-1-4419-6494-6_1} {\emph
  {\bibinfo {booktitle} {Rheology of Complex Fluids}}},\ \bibinfo {editor}
  {edited by\ \bibinfo {editor} {\bibfnamefont {J.~M.}\ \bibnamefont
  {Krishnan}}, \bibinfo {editor} {\bibfnamefont {A.~P.}\ \bibnamefont
  {Deshpande}}, \ and\ \bibinfo {editor} {\bibfnamefont {P.~B.~S.}\
  \bibnamefont {Kumar}}}\ (\bibinfo  {publisher} {Springer},\ \bibinfo
  {address} {New York},\ \bibinfo {year} {2010})\ pp.\ \bibinfo {pages}
  {3--34}\BibitemShut {NoStop}%
\bibitem [{\citenamefont {Warnez}\ and\ \citenamefont
  {Johnsen}(2015)}]{Warnez2015}%
  \BibitemOpen
  \bibfield  {author} {\bibinfo {author} {\bibfnamefont {M.~T.}\ \bibnamefont
  {Warnez}}\ and\ \bibinfo {author} {\bibfnamefont {E.}~\bibnamefont
  {Johnsen}},\ }\bibfield  {title} {\enquote {\bibinfo {title} {Numerical
  modeling of bubble dynamics in viscoelastic media with relaxation},}\ }\href
  {\doibase 10.1063/1.4922598} {\bibfield  {journal} {\bibinfo  {journal}
  {Phys. Fluids}\ }\textbf {\bibinfo {volume} {27}},\ \bibinfo {pages} {063103}
  (\bibinfo {year} {2015})}\BibitemShut {NoStop}%
\bibitem [{\citenamefont {Pearson}\ and\ \citenamefont
  {Middleman}(1977)}]{Pearson1977}%
  \BibitemOpen
  \bibfield  {author} {\bibinfo {author} {\bibfnamefont {G.}~\bibnamefont
  {Pearson}}\ and\ \bibinfo {author} {\bibfnamefont {S.}~\bibnamefont
  {Middleman}},\ }\bibfield  {title} {\enquote {\bibinfo {title} {Elongational
  flow behavior of viscoelastic liquids: Part {I}. modeling of bubble
  collapse},}\ }\href {\doibase 10.1002/aic.690230513} {\bibfield  {journal}
  {\bibinfo  {journal} {AIChE Journal}\ }\textbf {\bibinfo {volume} {23}},\
  \bibinfo {pages} {714--722} (\bibinfo {year} {1977})}\BibitemShut {NoStop}%
\bibitem [{\citenamefont {Jones}, \citenamefont {Walters},\ and\ \citenamefont
  {Williams}(1987)}]{Jones1987}%
  \BibitemOpen
  \bibfield  {author} {\bibinfo {author} {\bibfnamefont {D.~M.}\ \bibnamefont
  {Jones}}, \bibinfo {author} {\bibfnamefont {K.}~\bibnamefont {Walters}}, \
  and\ \bibinfo {author} {\bibfnamefont {P.~R.}\ \bibnamefont {Williams}},\
  }\bibfield  {title} {\enquote {\bibinfo {title} {On the extensional viscosity
  of mobile polymer solutions},}\ }\href {\doibase 10.1007/BF01332680}
  {\bibfield  {journal} {\bibinfo  {journal} {Rheol. Acta}\ }\textbf {\bibinfo
  {volume} {26}},\ \bibinfo {pages} {20--30} (\bibinfo {year}
  {1987})}\BibitemShut {NoStop}%
\bibitem [{\citenamefont {Day}\ \emph {et~al.}(2008)\citenamefont {Day},
  \citenamefont {Blanchard}, \citenamefont {English}, \citenamefont {Dobbie},
  \citenamefont {Williams}, \citenamefont {Garvey},\ and\ \citenamefont
  {Wong}}]{Day2008}%
  \BibitemOpen
  \bibfield  {author} {\bibinfo {author} {\bibfnamefont {M.}~\bibnamefont
  {Day}}, \bibinfo {author} {\bibfnamefont {R.}~\bibnamefont {Blanchard}},
  \bibinfo {author} {\bibfnamefont {R.}~\bibnamefont {English}}, \bibinfo
  {author} {\bibfnamefont {T.}~\bibnamefont {Dobbie}}, \bibinfo {author}
  {\bibfnamefont {R.}~\bibnamefont {Williams}}, \bibinfo {author}
  {\bibfnamefont {M.}~\bibnamefont {Garvey}}, \ and\ \bibinfo {author}
  {\bibfnamefont {D.}~\bibnamefont {Wong}},\ }\bibfield  {title} {\enquote
  {\bibinfo {title} {Shear and extensional rheometry of pdms tamponade agents
  used in vitroretinal surgery},}\ }\href {\doibase 10.1063/1.2964592}
  {\bibfield  {journal} {\bibinfo  {journal} {AIP Conf. Proc.}\ }\textbf
  {\bibinfo {volume} {1027}},\ \bibinfo {pages} {1411--1413} (\bibinfo {year}
  {2008})}\BibitemShut {NoStop}%
\bibitem [{\citenamefont {Williams}\ \emph {et~al.}(2010)\citenamefont
  {Williams}, \citenamefont {Day}, \citenamefont {Gavey}, \citenamefont
  {English},\ and\ \citenamefont {Wong}}]{Williams2010}%
  \BibitemOpen
  \bibfield  {author} {\bibinfo {author} {\bibfnamefont {R.~L.}\ \bibnamefont
  {Williams}}, \bibinfo {author} {\bibfnamefont {M.}~\bibnamefont {Day}},
  \bibinfo {author} {\bibfnamefont {M.~J.}\ \bibnamefont {Gavey}}, \bibinfo
  {author} {\bibfnamefont {R.}~\bibnamefont {English}}, \ and\ \bibinfo
  {author} {\bibfnamefont {D.}~\bibnamefont {Wong}},\ }\bibfield  {title}
  {\enquote {\bibinfo {title} {Increasing the extensional viscosity of silicone
  oil reduces the tendency for emulsification},}\ }\href {\doibase
  10.1097/IAE.0b013e3181babe0c} {\bibfield  {journal} {\bibinfo  {journal}
  {Retina}\ }\textbf {\bibinfo {volume} {30}},\ \bibinfo {pages} {300--304}
  (\bibinfo {year} {2010})}\BibitemShut {NoStop}%
\end{thebibliography}%

\end{document}